\documentclass[twocolumn]{autart}    

\usepackage{graphicx}          
\usepackage{url}
\usepackage[pdftex]{color}
\usepackage{amsmath}
\usepackage{amsfonts}
\usepackage{hyperref}
\usepackage{amssymb}
\usepackage{bbold}
\usepackage{tikz}
\usepackage{siunitx}
\usepackage{tabularray}
\usepackage{tabularx}
\usepackage{multirow}

\usepackage{algorithm}
\usepackage{algpseudocode}

\newtheorem{theorem}{Theorem}[section]
\newtheorem{corr}{Corollary}[section]

\newtheorem{lemma}{Lemma}[section]

\newtheorem{remark}{Remark}
\newtheorem{ass}{Assumption}
\usepackage{soul}

\newcommand{\R}{\mathbf{R}}
\newcommand{\Sb}{\mathbf{S}}

\newcommand{\N}{\mathbf{N}}

\newcommand{\E}{\mathbf{E}}

\newcommand{\Rc}{\mathcal{R}}
\newcommand{\Bc}{\mathcal{B}}

\newcommand{\Uc}{\mathcal{U}}
\newcommand{\Nc}{\mathcal{N}}

\newcommand{\Pc}{\mathcal{P}}

\newcommand{\tp}{^\top}
\newcommand{\inv}{^{-1}}
\newcommand{\beq}{\begin{equation}}
	\newcommand{\eeq}{\end{equation}}
\newcommand{\bmat}{\begin{bmatrix}}
	\newcommand{\emat}{\end{bmatrix}}

\DeclareMathOperator{\diag}{diag}

\newcommand{\proof}{\noindent {\it Proof. }}
\newcommand\m[1]{\begin{bmatrix}#1\end{bmatrix}}
\DeclareMathOperator{\eig}{eig}
\DeclareMathOperator{\argmin}{arg\,min}

\definecolor{purple}{rgb}{0.62, 0.0, 0.77}

\definecolor{Royalblue}{cmyk}{1,0.30,0.2,0.2}

\definecolor{Indigo}{rgb}{0.1, 0.35, 0.62}

\definecolor{cobalt}{rgb}{0.0, 0.28, 0.67}

\newcommand{\rev}{}

\begin{document}

\begin{frontmatter}

\title{Distributionally Robust LQG {\rev Control} under Distributed Uncertainty \thanksref{footnoteinfo}} 

\thanks[footnoteinfo]{Lucia Falconi’s activity is supported by Fondazione CARIPARO, Italy (Borse di Dottorato CARIPARO 2020)}

\author[eth]{Lucia Falconi}\ead{lfalconi@ethz.ch},    
\author[unipd]{Augusto Ferrante}\ead{augusto@dei.unipd.it},               
\author[unipd]{Mattia Zorzi }\ead{zorzimat@dei.unipd.it}  

\address[eth]{Automatic Control Laboratory, ETH Z\"{u}rich, Z\"{u}rich, Switzerland}  
\address[unipd]{Department of Information Engineering, University of Padova, Padova, Italy}

\begin{keyword}                           
Optimal control; Robustness; Stochastic systems; Dynamic programming; Relative entropy.             
\end{keyword}                             

\begin{abstract}                          
A new paradigm is proposed for the robustification of the LQG controller against distributional uncertainties on the  noise process. 
Our controller optimizes the closed-loop performance in the worst possible scenario
 under the constraint that the noise distributional aberrance
does not exceed a certain threshold limiting the 
{\rev relative entropy}
between the actual noise  distribution and the nominal one.  
The main novelty is that the bounds on the distributional aberrance can be arbitrarily distributed along the whole disturbance trajectory.
This is a problem for which, notwithstanding significant attention
given in the recent literature, so far only relaxed or approximated solutions
have been derived. We denote this Distributed uncertainty Distributionally robust LQG  problem with the acronym D$^2$-LQG.
\end{abstract}

\end{frontmatter}

\section{Introduction}
Linear Quadratic (LQ) optimal control is arguably one of the most fundamental ideas
in modern control theory with ramifications in virtually all the fields of the discipline.
In particular, this is a natural approach for the control of  multiple-input and multiple-output (MIMO) systems.
The theory may be easily extended to stochastic systems by considering the Linear Quadratic Gaussian (LQG) control which is the core idea in the
development of many advanced control techniques, such as all the MPC and MPC-like methods \cite{rawlings2017model}.

The LQG method involves the design of a
controller which optimizes the performance based on a nominal
model of the process which is affected by a Gaussian distributed disturbance with known mean and covariance. 
Thus, in the nominal situation the LQG controller guarantees optimal performance.
However, soon after its introduction, it was recognized \cite{doyle1978} that, as for a kind of waterbed effect, optimality makes the LQG controller vulnerable to model uncertainties.
This lack of robustness has motivated a formidable stream of research aiming
to ensure that the controller provides satisfactory performance even in the presence of model misspecifications and inaccuracies.
The extensive research conducted in this area has resulted in a multitude of available approaches, including $H^\infty$ control methods, based on the minimization of $H^\infty $ performance criteria (see for example \cite{ZHOU1985,francis1987course,doyle1989,BERGELING2020}), and risk-sensitive LQG control, where the standard quadratic loss function is replaced by an exponential quadratic function \cite{1973jacobson,Whittle}.
An alternative and very attractive approach that has 
recently received significant attention is  \emph{Distributionally robust control (DRC)}, see e.g. \cite{yang2020wasserstein,dixit2022distributionally,fochesato2022data,hakobyan2022wasserstein}. Such paradigm takes  into account  uncertainty  in the probability distribution underlying a stochastic system.
Instead of assuming a given distribution, DRC methods design a control policy which minimizes a given cost under the worst-case distribution in the so-called \emph{ambiguity set}. 

Several types of ambiguity sets have been employed in DRC, exploiting for example moment constraints \cite{van2015distributionally}, total variational distance \cite{tzortzis2015dynamic,dixit2022distributionally}, 
Wasserstein distance \cite{yang2020wasserstein,fochesato2022data,hakobyan2022wasserstein,kim2020minimax}. 
In the seminal paper  \cite{Petersen2000}, the ambiguity set is a ball  defined in the relative-entropy topology and centered at the nominal distribution. The latter  has several attractive features. 
First and foremost relative entropy is the natural ``metric'' between systems when they are identified from data, \cite{hansen2008robustness,ZORZI2017133}. 
Moreover, relative entropy is a prominent metric in information theory
and has important structural properties.

The DRC problem in \cite{Petersen2000} limits  the distribution of the uncertainty along the time interval using a {\em single} constraint on the overall noise distribution on the entire  time interval. This  is the natural way to model the uncertainties when the discrepancy between the nominal and the actual system is due to the action of an adversary who can manage a limited mismatch budget to perturb the nominal system and, when convenient, is allowed to concentrate most (or all) such a budget in a few time points, or even in a single one.  However, in most practical situations there is not a real adversary and
the model mismatch is a consequence of modelling approximations and random fluctuations. 
In these cases, allowing the concentration of the uncertainty on few time steps
may lead to unrealistic scenarios and to overly conservative conclusions
because the same effort is usually made to model each time step. Accordingly, a more realistic approach is to ``distribute the uncertainty'', by imposing 
a constraint to each time step; a similar  point of view has been adopted in \cite{levy2013,ZorziTAC2017,9521824,abadeh2018wasserstein} for the robust filtering  setting.
We clarify the practical need of a robust DRC paradigm with  distributed uncertainty with two examples:\\
1. Assume that 
we want to design a robust LQG optimal controller for an airplane.
The Gaussian noise models the effects of the wind which, in practice, is not Gaussian.
Therefore, we need to guarantee robustness against different distributions of the 
wind disturbance. By adopting  a single constraint as in \cite{Petersen2000}, it may happen that the
worst-case distribution differs from the nominal one at just one time interval
but in that time interval the difference is unreasonably large:
the controller must be able to counteract a totally unrealistic wind concentrated at a single time point with drammatic degradation of the performance.
The possibility of distributing the uncertainty along the whole trajectory and to 
select different uncertainties radii depending on the reliability of the wind forecasts at each point is clearly an interesting feature in this scenario.\\
2. Consider the case when the difference between the nominal Gaussian  noise and the actual  one is used to account for errors in the model parameters.
Also in this case, the possibility of distributing the uncertainty along the whole trajectory is clearly advantageous with respect to a potential worst-case where the model mismatch is unreasonably concentrated at a single time interval.

For these reasons, a significant stream of literature has been recently devoted 
to distribute the uncertainty along the whole time horizon.
This problem appears however quite challenging and so far 
the solution has only been derived for relaxed or approximated versions of the problem. 
In \cite{yang2020wasserstein} and  \cite{tzortzis2015dynamic}, the control law is obtained 
by means of the value iteration and the policy iteration method, respectively.
Another strategy of approximation is to relax the constraints at each time step, i.e. the constraints are replaced by penalty terms in the objective function in the DRC problem, \cite{yang2020wasserstein,hakobyan2022wasserstein}. In conclusion, it is apparent that a complete theory for this
problem cannot be obtained as a direct extension or simple adaptation 
of the single constraint counterpart of \cite{Petersen2000}.

The aim of this paper is to derive a complete theory 
addressing  the exact solution of the DRC problem where the distributed uncertainty is described at each time step by an  ambiguity set defined according to the relative entropy.  We denote this Distributed uncertainty Distributionally robust LQG  problem with the acronym D$^2$-LQG, {\rev where $D^2$ stands for Distributed uncertainty Distributionally robust.}


 {\rev 
 The  main idea for our theory  is to combine  a dynamic programming technique with a generalization of Von Neumann's minimax Theorem.  Since this theorem cannot be applied directly due to the non-compactness of both the maximization and minimization sets,  we establish a preliminary theorem that allows us to restrict  our attention to compact subsets. This  approach leads to a new problem which can in turn be attacked employing the Lagrange duality theory and the duality between free-energy and relative entropy. 
In this way, we arrive to a new high dimensional, yet tractable, optimization problem for which we propose an effective numerical algorithm.
The overall solution takes the form of a risk-sensitive controller with a time-varying risk-sensitive parameter.
The structure of the resulting D$^2$-LQG controller does resemble the one of the standard linear quadratic regulator obtained in absence of noise and uncertainty. While the structure of the controllers may appear similar, the arguments needed to derive the solution in the D$^2$-LQG case are much more complex. Indeed, whereas in the linear quadratic regulator setting the solution is simply obtained by completing the squares, the derivation of the D$^2$-LQG controller, as discussed above, is significantly more complex, reflecting the challenging nature of the problem.
}

The outline of the paper is as follows. 
In Section \ref{sec:pb} 
we formulate the D$^2$-LQG control problem with relative-entropy ambiguity sets.
Section \ref{sec:worst-perf-analysis} deals with a
detailed analysis complete with all the proofs of the
worst-case performance problem with respect to a quadratic cost:
a closed-form solution is derived as a preliminary step towards the solution of our problem.
Some of the preliminary results established in Section \ref{sec:worst-perf-analysis} were presented without proof in our preliminary CDC paper \cite{Falconi_CDC22}. In Section \ref{sec:minmax},
we derive the solution of the D$^2$-LQG control problem.  
 An illustrative example is presented in Section \ref{sec:simulation} which shows the advantages of distributing the model uncertainty to trade off optimality and robustness.
An extension to the previous theory is proposed in Section \ref{sec:e2}, by letting the ambiguity set radius to depend on the norm of the state and control variables.   
Finally, conclusions and future works are discussed in Section \ref{sec:conc}.
We extensively use some relative entropy properties and convex optimization results that, for the convenience of the reader, are summurized in  Appendix \ref{sec:app_KL} and \ref{sec:app_opt}, respectively.

\subsection*{Notation}
$\R$ is the real line; $\R_{+}$ ($\R_{++}$) is the set of nonnegative (positive) scalars.
{\rev  The Cartesian product over $n$ sets $X_1, \dots , X_n$ is denoted by $X_1 \times \cdots \times X_n.$ }
Given a matrix $A$, $A\tp$ denotes the transpose. If $A$ is a square matrix, $|A|$ denotes its determinant and, if it is also symmetric, we write $A \succ 0$ to denote that $A$ is positive definite and   $A \succeq 0$ to denote that  $A$ is positive semi-definite. 
 We denote by $\eig (A)$  the set of eigenvalues of the matrix $A.$
We denote by $\Sb^n$, $\Sb_{+}^n$, and $\Sb_{++}^n$ the set of
$n\times n$ symmetric matrices, the set of $n\times n$ positive semidefinite matrices and the set of $n\times n$ positive definite matrices, respectively.
$I$ is the identity matrix. 
{\rev We denote as $\mathcal{P}(\R^n)$ the set of probability measures taking values on $\R^n.$
To ease the exposition, and with some abuse of notation, $x$ denotes both a random vector and its realization; moreover, $f(x)$ denotes the probability density function  of such random vector evaluated at $x$. }
$ x \sim  \Nc(\mu;\Sigma)$ means that $x$ is a Gaussian random
variable with mean $\mu$ and covariance matrix $\Sigma$, whereas $ x \sim  \Uc(a,b) $  means that $x$ is a random variable  uniformly distributed over the interval $[a,b].$  
The symbol $\E [\cdot]$ denotes the expectation with respect to the probability density function $f(\cdot)$, whereas  $\tilde \E [\cdot]$ denotes the expectation with respect to the density function $\tilde f(\cdot).$ 
Given two probability measures $\tilde f$ and $f$ defined on $\R^n$, we denote by 
$$
{\rev \Rc(\tilde f || f )} := 
\int_{\R^n} \ln \frac{\tilde f(x)}{f(x)} \tilde f(x) dx
$$
the \emph{relative entropy} (or \emph{Kullback-Leibler} divergence)
between $\tilde f$ and $f$.

\section{Problem formulation}\label{sec:pb}

Consider a discrete-time linear stochastic systems
\beq \label{eq:refsys}
x_{t+1} = Ax_t + Bu_t + v_t  \qquad t=0,\dots,N
\eeq
where $x_t \in \R^n$ is the state vector, $u_t \in \R^m$ is the control input and $ v_t \in \R^n$ is a random vector representing the system disturbance. The initial state $x_0 = \bar x_0 $ is  a deterministic quantity.
{\rev The noise $v_t$ follows a probability distribution  that is \emph{unknown} to the  control designer. The control designer has only access to a \emph{nominal} probability distribution $f_t,$ which is assumed to be a Gaussian density function with zero mean and covariance $V  \succ 0.$ 
This nominal distribution $f_t$ (as well as the parameters of the nominal model) is typically estimated from data using system identification methods \cite{lennart1999system} or dictated by prior knowledge of the system.
However, it is usually subject to modelling errors. 
To account for these errors, we introduce a \emph{perturbed} probability density function $\tilde f_t.$ }
We denote by $\tilde f$ the cumulative joint distribution of the whole noise trajectory
$v=[v_0,\dots,v_N]$ and by $f$ its nominal value.
The matrices $A, B$ and $V$ have appropriate dimensions, and, for the ease of notation, they are assumed time-independent. 
Generalization to time-varying systems only entails heavier notations. The conclusions however are exactly the same and can be reached by the exact same arguments.

{\rev Our goal is to design an optimal controller that is robust against deviations of $\tilde f_t$ from $f_t.$
To model a limit on the admissible perturbed noise distributions $\tilde f_t,$ 
we use a relative entropy
tolerance.
} 
Specifically, for each time instant $t$, given $x_t \in \R^n,$ we define the \emph{ambiguity set} as 
the convex ball of functions
\beq  \label{eq:amb_set}
\Bc_t (x_t) := \big \{ \tilde f_t \in  \Pc (\R^n) : {\rev \Rc( \tilde f_t || f_t)} \leq d_t + \frac{1}{2} \Vert z_t \Vert^2   \big\}. \eeq
Here, the tolerance $d_t>0$ quantifies the constant mismatch budget allowed at time $t,$  whereas the signal 
$$z_t := E_1 x_t,$$ 
with $E_1 \in \R^{p \times n},$
defines the mismatch budget proportional to the norm of the state.
Whenever convenient, we will use the shorthand ${\rev \Rc_t}$ to denote $ {\rev \Rc( \tilde f_t || f_t)} .$

The performance index is the standard symmetric quadratic function of state and control action  for stochastic systems:
\begin{align} \label{eq:cost}
\begin{aligned}J(u,\tilde{f},\bar x_0) = \tilde \E  
\big[ & \sum_{t=0}^{N} \frac{1}{2}  (x_t\tp Q x_t + u_t\tp R u_t) + \\ 
& \frac{1}{2}  x_{N+1}\tp Q_{N+1} x_{N+1} \big | x_0 = \bar x_0 \big],
\end{aligned} \end{align}
where $ Q \succeq 0$, $ Q_{N+1} \succ 0,$  $R \succ 0$ and $u = (u_0, \dots , u_N).$ 
We assume that the pair $(A,Q)$ is observable.

The admissible control policies are state feedback of the form
\beq \label{eq:contrl} 
u_t = \pi_t(x_t), \quad  t = 0, \dots, N 
\eeq
where $\pi_t: \R^n \to \R^m$ is a measurable function. 
{\rev We assume that the admissible noise distribution  $\tilde f_t$ belongs to the ambiguity set \eqref{eq:amb_set} and that it is a measurable function of the state-action pair. Denoting this function by $\gamma_t,$ the noise distribution can be expressed as 
\beq  \label{eq:noise-dist}
\tilde f_t = \gamma_t(x_t, u_t) \in \Bc_t (x_t) \qquad t=0, \dots, N
\eeq
with $\gamma_t: \R^{n+m} \to {\rev \Pc (\R^n) }$ measurable.
Therefore, $ \pi_t$ is the control policy and  $\gamma_t$ is the adversary policy 
(see Remark \ref{rem:DRCgame} for the interpretation of the D$^2$-LQG control problem as a two-players zero-sum game).
}
We address
the following minimax optimization problem:
\beq \label{pb:DRC} 
\inf_{\pi \in \Pi} \sup_{\gamma \in \Gamma} J(\pi,\gamma,\bar x_0)  
\eeq 
where   
{\rev 
$ \Pi := \{ \pi = (\pi_0, \dots, \pi_N) \hbox{ s.t. } \pi_t: \R^n \to \R^m \}$
is the set of admissible control policies and 
$\Gamma := \{ \gamma = (\gamma_0, \dots \gamma_N)
\hbox{ s.t. } \gamma_t: \R^{n+m} \to {\rev \Pc (\R^n) }, \; \gamma_t (x_t, u_t) \in  \Bc_t (x_t) \; \forall (x_t, u_t)\}$
is the set of admissible adversary policies.
\footnote{\rev The notations $J(u,\tilde{f},\bar x_0)$ and $J(\pi,\gamma,\bar x_0) $ are used interchangeably in the paper: $J(u,\tilde{f},\bar x_0)$ emphasizes the dependence of the cost on the specific input $u$ and the perturbed density $\tilde f.$ Conversely, $J(\pi,\gamma,\bar x_0) $  highlights the role of the control policy $\pi$ and the adversary policy $\gamma$ in determining $u$ and $\tilde f$, respectively.  This is mathematically consistent  in view of the 
relations  (\ref{eq:contrl}) 
and \eqref{eq:noise-dist} providing the links between $\pi$ and $u$, and between $\gamma$ and and $\tilde f$, respectively.
} }
Our aim is to prove that $\inf$ and $\sup$ in (\ref{pb:DRC})
are indeed attained and to design the optimal control policy 
$u_t^\star = \pi_t^\star(x_t)$.

\begin{remark} \label{rem:DRCgame}
A suggestive interpretation of Problem (\ref{pb:DRC}) is the following:
we set up a two-players zero-sum game in which Player I is the controller, and Player II is an hypothetical adversary.
Player II selects the adversary policy $\gamma_t$ in order to maximize the cost, and Player I determines the control policy $\pi_t$ to contain the damage by minimizing the same cost $J$ against the best effort of the adversary.
In this way the optimal control policy is robust 
against the worst possible scenario.
\end{remark}

\section{Worst performance analysis}\label{sec:worst-perf-analysis}
In this section, we consider the {\em absence of a control input} i.e. the case in which $u_t \equiv 0$; in this simple case we characterize the worst-case performance with respect to a quadratic cost function for our stochastic uncertain system.
{\rev The solution to this problem is a crucial preliminary step towards the derivation of the D$^2$-LQG controller  in the next section. 
Indeed, by solving the worst performance analysis problem, we identify the worst model within the ambiguity set, 
which the D$^2$-LQG controller must counteract. 
This analysis also provides a  valuable insight into the nature of the worst-case model.  
Additionally, tackling this simpler problem enhances the comprehension for the analysis in Section 4, where  a similar, yet more sophisticated, reasoning is applied.
}

By assuming that the input signal is identically equal to zero, the system state equation \eqref{eq:refsys} becomes
\beq \label{eq:WP:refsys}
x_{t+1} = Ax_t + v_t,  \quad t=0,\dots,N
\eeq 
and the cost function becomes 
$$
\mathcal{J} (\tilde{f},\bar x_0) = \tilde \E \big[ \frac{1}{2} \sum_{t=0}^{N}  x_t\tp Q x_t + \frac{1}{2} x_{N+1}\tp Q_{N+1} x_{N+1}| x_0 = \bar x_0 \big]
$$
where $ Q \succeq 0$, $ Q_{N+1} \succ 0.$ 
This is the typical cost function in linear quadratic problems for stochastic autonomous systems.
To characterize the worst performance, we adopt a hypothetical opponent that selects at each time instant $t$ the probability distribution $\tilde f_t $ in an adversarial way. 
The problem under consideration is
\beq \label{pb:WP}
\sup_{\gamma \in \Gamma} \mathcal{J} (\gamma,\bar{x}_0)
\eeq 
where 
$\Gamma := \{ \gamma = (\gamma_0, \dots \gamma_N) \; : \; \gamma_t(x_t) = \tilde f_t \in \Bc_t (x_t)  \}$
is the set of adversary policy space.
{\rev Note that, $\mathcal{J} (\tilde{f},\bar{x}_0)=\mathcal{J} (\gamma,\bar{x}_0)$ and the notations are used interchangeably in the paper. }

\begin{ass}\label{ass:obs}
The  pair $(A,Q)$ is observable.
\end{ass}

Exploiting the constraints' and the objective function's special structure, it is possible to obtain the overall solution by optimizing a sequence of $N+1$ single-variable constrained optimization problems by adopting a \emph{backward dynamic programming} technique \cite[ch. I]{rawlings2017model}.
Let $\mathcal{V}_{t}: \R^n \to \R$ be the value function defined as
\begin{align*}
  \mathcal{V}_{t} (x) := \sup_{\gamma \in \Gamma} \tilde \E \frac{1}{2} 
  \big[  \sum_{s=t}^{N} x_s\tp Q x_s  + x_{N+1}\tp Q_{N+1} x_{N+1}  | x_t = x \big].
\end{align*} 
Clearly, $\mathcal{V}_t (x)$ represents the worst-case expected cost-to-go from stage $t$ given $x_t = x.$
The value function satisfies the recursion
\beq\label{eq:wp_dprecurs}
\mathcal{V}_t (x)  =  \sup_{\tilde f_t \in \Bc_t(x) } \tilde \E \Big[ \frac{1}{2} x_t \tp Q x_t + \mathcal{V}_{t+1} ( Ax_t + v_t) | x_t = x\Big] 
\eeq
for $t=0, \dots, N.$
For $t = N+1 $ we have that 
$$ \mathcal{V}_{N+1} (x) =   \frac{1}{2} x\tp Q_{N+1} x. $$
Then, according to the dynamic programming principle 
$$ \sup_{\gamma \in \Gamma} {\mathcal J}(\gamma,\bar x_0) = \mathcal{V}_0 (\bar{x}_0).$$

\begin{theorem}\label{th:wp_main}	
The value function of Problem \eqref{pb:WP} 
has the following form
\beq \label{eq:wp_value_fcn} 
\mathcal{V}_{t} (x) = \min_{\omega_{t} \in \Omega_t , \dots, \omega_{N} \in \Omega_{N} } \mathcal{W}_{t} (x, \omega_{t} , \dots, \omega_{N} )  \eeq
 where  
\begin{align}  \label{eq:w_function} \begin{aligned}
\mathcal{W}_{t} (x, \omega_{t} , \dots, \omega_{N}) = \frac{1}{2} x \tp {\rev S_t} x + \sum_{s=t}^{N} r_s 
\end{aligned} \end{align} 
Here, 
\beq  \label{eq:Lambda_set}
\Omega_t := \big\{ 
\omega_t \in \R_{++} \; : \; \omega_t > \max \{\eig({\rev S_{t+1}} V) \} \big\}
\eeq
where ${\rev S_t} \in \mathbf{S}^n_{++}$ is recursively computed as 
\beq\label{eq:wp_Riccati}  
{\rev S_t}  =  Q + \omega_{t} E_1\tp E_1 +  A\tp \left({\rev S_{t+1}}^{-1} -  \frac{V}{\omega_{t}} \right)^{-1} A 
\eeq 
with terminal conditions ${\rev S_{N+1} }= Q_{N+1},$
and 
\beq \label{eq_wp_z}
r_t =  -  \frac{\omega_t}{2} \ln \left| I - \frac{ {\rev S_{t+1}} V}{\omega_t} \right|  {\rev +} \omega_t d_t  .
\eeq
{\rev 
Moreover, the minimum in \eqref{eq:wp_value_fcn} is  attained, meaning that there exists some $(\omega_{t}^\star, .. , \omega_{N}^\star) \in \Omega_t \times \dots \times \Omega_{N}$
such that $\mathcal{V}_{t} (x)  = \mathcal{W}_{t} (x, \omega^\star_{t} , \dots, \omega^\star_{N} ).$
}
\end{theorem}

Before proving Theorem \ref{th:wp_main} theorem, we state some preliminary results.

\begin{lemma}\label{lemm:pd}
    Let ${\rev S_{t+1}} \in \Sb_{++}^n $ and $\omega_t \in \Omega_t.$ Assume that Assumption \ref{ass:obs} is satisfied.
    Then, ${\rev S_{t+1}}\inv - V/\omega_t \succ 0 $ and ${\rev S_t}$ defined by \eqref{eq:wp_Riccati} is positive definite.
\end{lemma}
\proof
From ${\rev S_{t+1}} \succ 0$ and the definition of the set $\Omega_t,$ we immediately conclude that ${\rev S_{t+1}}\inv - V/\omega_t \succ 0.$
Then, $ {\rev S_t} \succeq 0$ as it is the sum of three positive-semidefinite matrices.
It remains to show that ${\rev S_t}$ is not singular. 
We prove this point by contradiction. 
Assume that there exists $y\neq 0$ such that $y\tp {\rev S_t} y = 0,$ 
{  \rev 
that is
\beq \label{eq:proof_lemma}
y\tp Q y + y\tp \omega_{t} E_1\tp E_1 y +  y\tp A\tp \left({\rev S_{t+1}}^{-1} -  \frac{V}{\omega_{t}} \right)^{-1} A y = 0.
\eeq
Since each term on the left-hand side of \eqref{eq:proof_lemma} is non-negative, condition \eqref{eq:proof_lemma} implies that each term must individually be equal to 0.
Therefore we have $Qy = 0$ and 
$({\rev S_{t+1}}^{-1} -  \frac{V}{\omega_{t}} )^{-1} A y = 0. $
Given that $({\rev S_{t+1}}^{-1} -  \frac{V}{\omega_{t}} )^{-1}$ is invertible, the second equation implies that $Ay = 0.$
Thus, we have found a non-zero vector $y$ such that
$ 	Q y = 0 $
and $ A y = 0.$ 
This contradicts Assumption \ref{ass:obs} as $y \neq 0$ belongs to the non observable subspace of the pair $(A, Q).$ }
$\hfill \qed$

\begin{theorem} \label{th:wp_strongdual}
  Given $x \in \R^n$ and ${\rev S_{t+1}} \in \Sb_{++}^n $, let
\begin{align}
\mu & = \sup_{\tilde f_t \in \Bc_t(x)} \tilde \E \left[ \frac{1}{2} (Ax + v_t)\tp {\rev S_{t+1}} (Ax + v_t) \right] \label{pb:wc_primal} \\
\phi & = \min_{\omega_t \in \Omega_t}   \frac {1}{2} x\tp ({\rev S_t} - Q ) x\tp + {\rev r_t}   \label{pb:wc_dual}
\end{align}
with $\Omega_t,$ ${\rev S_t} $ and $\rev r_t $ defined by  \eqref{eq:Lambda_set}-\eqref{eq_wp_z} respectively.
Then $\mu < \infty,$ $\mu = \phi $ and the minimum in \eqref{pb:wc_dual} is achieved for some $\omega_t^\star \in \Omega_t$ . 
\end{theorem}

\proof 
We exploit the Lagrange duality theory to prove this theorem.
First, we show that $\mu < \infty.$
Let $\omega_t \geq 0,$ and  
$L(\tilde f_t, \omega_t) := \tilde \E [ \frac{1}{2} (Ax + v_t)\tp {\rev S_{t+1}} (Ax + v_t) ]  - \omega_t ( {\rev \Rc_t} - \frac{1}{2} \Vert E_1 x \Vert^2 - d_t ) 
$
be the Lagrangian for Problem \eqref{pb:wc_primal}.
The dual function is defined as $ D (\omega_t) := \sup_{\tilde f_t \in \Pc(\R^n)} L (\tilde f_t, \omega_t). $
It is straightforward to verify that 
$ \mu \leq D (\omega_t) $
for every $\omega_t \geq 0.$
Next, we show that there exists $\omega_t \in \R_+$ such that  
$D (\omega_t) < \infty$ and, consequently, $\mu  < \infty.$
To this end, we first notice that we can rule out the case in which $\omega_t = 0.$
Indeed, if $\omega_t = 0$ we have that 
 $ D(\omega_t) = \sup_{\tilde f_t \in \Pc(\R^n)} \tilde \E \left[ \frac{1}{2} (Ax + v_t)\tp {\rev S_{t+1}} (Ax + v_t) \right], $
 which is clearly unbounded since ${\rev S_{t+1}}$ is  strictly positive. 
Then, we evaluate the function $ D (\omega_t)  $ for $\omega_t > 0.$  We obtain that the dual function
\begin{align*} 
& D (\omega_t)  =  \frac{\omega_{t}}{2} \Vert E_1 x \Vert^2  + \omega_{t} d_t  +\omega_t \times \\
& \big( \sup_{\tilde f_  \in \Pc(\R^n) }    \int_{\R^n}  (Ax + v_t) \tp \frac{{\rev S_{t+1}}}{2\omega_{t}} (Ax + v_t) \tilde f_t (v_t) dv_t  -  {\rev \Rc_t} \big),
\end{align*}
{\rev where we recall that $v_t$ is used to denote both the random variable and its realization as the argument of $\tilde f_t$. }
From Lemma \ref{lemma:KLopt}, it follows that 
\begin{align*}
 &\sup_{\tilde f_  \in \Pc(\R^n) }    \int_{\R^n}  (Ax + v_t) \tp \frac{{\rev S_{t+1}}}{2\omega_{t}} (Ax + v_t) \tilde f_t (v_t) dv_t  - {\rev \Rc_t}\\ 
& =  \ln \Big(  \int_{\R^n}  \exp \{ (Ax + v_t) \tp \frac{{\rev S_{t+1}}}{2\omega_{t}} (Ax + v_t) \} f_t (v_t) dv_t  \Big).
\end{align*}
In view of Lemma \ref{lemma:GaussIntegral}, the integral in the latter equation converges if and only if ${\rev S_{t+1}}\inv - V/\omega_t \succ 0,$ in which case
$$D (\omega_t) = \frac{1}{2} x\tp ({\rev S_t} - Q) x - \frac{\omega_t}{2} \ln |I - \frac{{\rev S_{t+1}}V}{\omega_t} |  + \omega_t d_t.$$
This concludes the first part of the theorem, as we have shown that there exists $\omega_t \in \R_+$ such that $D(\omega_t) < \infty.$ \\
We are now ready to prove that $\mu = \phi$ and the minimum in \eqref{pb:wc_dual} is attained.
To this end, we sensibly apply  Theorem \ref{lemma:luenberger} 
 with $X$ as the linear vector space of functions from $ \R^n$ to $\R$ and $\Omega$ the set of probability density functions on $\R^n$.  The objective function corresponds to  $\tilde \E [ (Ax + v_t)\tp {\rev S_{t+1}} (Ax + v_t) ]$ which is a linear, thus concave, map of $ \tilde f_t.$
The function $g(\cdot) $ corresponding to $ ({\rev \Rc_t}  - \frac{1}{2} \Vert E_1 x \Vert^2   - d_{t}) $ is a convex function of $  \tilde f_t$ thanks to the properties of the relative entropy. The supremum $\mu $ is finite as shown in the first part of the proof. 
{\rev Finally, the probability density function $\tilde f_t =  f_t  $ is strictly feasible. Indeed, from the properties of the Kullback-Leibler divergence (see Appendix \ref{sec:app_KL}), we have that  
$\mathcal{R}_t = \mathcal{R} (f_t || f_t) = 0.$
Therefore, since $d_t > 0,$ it follows that
$ 0 = \mathcal{R}_t  < d_t + \frac{1}{2} \Vert z_t \Vert^2.$
Thus, $\tilde f_t = f_t$ satisfies the inequality constraint with strict inequality, meaning that it is strictly feasible. 
}
Hence, the conditions of Theorem \ref{lemma:luenberger} are satisfied. 
Therefore, by exploiting the form of the dual function $D(\omega_t)$ derived in the first part of this proof, the thesis is immediatly obtained. 
$\hfill \qed$

\begin{lemma}\label{lemma:WP:conv}
Given $x \in \R^n,$ the function $\mathcal{W}_t (x, \omega_{t}, .., \omega_N) $ defined in \eqref{eq:w_function} is jointly convex in $(\omega_{t}, .., \omega_{N}) \in \Omega_t \times .. \times \Omega_N$ for any $t=0,\dots, N.$
\end{lemma}

\proof  
We show that 
\begin{align}\label{eq:proof}
\begin{aligned}
\mathcal{W}_t 
= &\sup_{ \tilde f_t, \dots , \tilde f_N \in  \Pc(\R^n)}    \tilde \E 	 \Big[  \sum_{k=t}^{N}  \frac{ x_k\tp Q x_k}{2}  +  \frac{x_{N+1}\tp  {\rev S_{N+1}}x_{N+1}}{2}    \\ 
& \qquad \qquad \quad
- \sum_{k=t}^{N} \omega_k ({\rev \Rc_k} -  \frac{1}{2} \Vert z_k \Vert ^ 2  - d_k) | x_t = x\Big] 
\end{aligned}
\end{align}
for $(\omega_{t}, .., \omega_{N}) \in \Omega_t \times .. \Omega_N.$  
Hence, since $\mathcal{W}_t$ is the pointwise maximum of a set of affine functions of $(\omega_{t}, \dots , \omega_{N}),$ it is convex by Lemma \ref{lemma:convex}. 
To prove \eqref{eq:proof}, consider the optimization problem in the right hand-side, which we denote by $p.$ 
We have that 
\begin{align*}
p &= \sup_{ \tilde f_t, \dots , \tilde f_{N-1}}    \tilde \E 	 	\Big[  \sum_{k=t}^{N-1}  \frac{x_k\tp Q x_k}{2}   - \sum_{k=t}^{N-1} \omega_k ({\rev \Rc_k} -  \frac{\Vert z_k \Vert ^ 2 }{2}  - d_k) 
\\
& - \omega_N(  \frac{1}{2} \Vert z_N \Vert ^ 2  - d_N )
+ \sup_{ \tilde f_N \in \Pc (\R^n) }
\tilde \E \big[  (Ax + v_N) \tp {\rev S_{N+1}}
\\
& \times  (Ax + v_N)
  - \omega_N {\rev \Rc_N} | x_N \big]  \Big]  .
\end{align*}
Applying first Lemma \ref{lemma:KLopt} and then Lemma \ref{lemma:GaussIntegral}, we obtain that 
\begin{align*}
&\sup_{ \tilde f_N }
\tilde \E \big[  (Ax_N + v_N)\tp {\rev S_{N+1}} (Ax_N + v_N) - \omega_N {\rev \Rc_N} | x_N \big] 
 \\
& = \omega_N \ln \big(  
\E \big[ e^{ (Ax_N + v_N)\tp \frac{Q_{N+1}}{2 \omega_N} (Ax_N + v_N) }  | x_N  \big] \big)  \\
&= x_N\tp ({\rev S_N} - E_1\tp E_1) x_N  -  
\frac{\omega_N}{2} \ln \left| I - \frac{{\rev S_{N+1}}V}{\omega_N} \right| . 
\end{align*}
Therefore, 
\begin{align*}
    p = &\sup_{ \tilde f_t, \dots , \tilde f_{N-1}}    \tilde \E \Big[  \sum_{k=t}^{N-1}  \frac{1}{2} x_k\tp Q x_k  + \frac{1}{2}x_N\tp {\rev S_N} x_N \\  
    &- \sum_{k=t}^{N-1} \omega_k ({\rev \Rc_k} -  \frac{\Vert z_k \Vert ^ 2 }{2}  - d_k)  | x_t = x\Big] + r_N.
\end{align*}
We now focus on the optimization with respect to $\tilde f_{N-1}.$ This problem is identical in structure to the previous one, hence we can repeat the previous arguments.
In conclusion, by recursively applying Lemma \ref{lemma:KLopt} and then Lemma \ref{lemma:GaussIntegral} we obtain \eqref{eq:proof}.
$\hfill \qed$

Finally, we are ready to prove Theorem \ref{th:wp_main}

\emph{Proof of Theorem \ref{th:wp_main}}
The theorem can be proved via mathematical induction.
At $ t = N+1,$ the statement is true as $V_{N+1} (x) = x\tp {\rev S_{N+1}} x.$
Now suppose that the induction hypothesis holds for $t+1.$
Then, it follows from \eqref{eq:wp_dprecurs} that 
\begin{align*} 
\mathcal{V}_t(x) = \sup_{\tilde f_t \in \Bc_t(x) } \; 
\min_{ \omega_{t+1} \in \Omega_{t+1}, ..., \omega_{N} \in \Omega_{N} }
\frac{1}{2} x\tp Q x_t + \\
\tilde \E[ \mathcal{W}_{t+1}  (Ax + v_t,\omega_{t+1} , .., \omega_{N} )  ] 
\end{align*}
Now, we apply  the  \emph{minimax  Sion's Theorem} \ref{lemma:minmax} in order to switch the maximization and minimization operation. To this end, notice that the objective function is linear (thus continuous and quasi-concave) in $\tilde f_t,$ and continuous and convex (thus quasi-convex) in $(\omega_{t+1}, \dots, \omega_{N} )$ in view of Theorem \ref{lemma:WP:conv}.
Moreover, $\Bc_t$ and $\Omega_{t+1} \times \dots \times \Omega_{N} $ are convex sets.
Finally, since a minimum point does exists, we can restrict the search of the minimum  to a compact convex subset of $\Omega_{t+1} \times \dots\times \Omega_{N}.$   
Hence, the conditions of Theorem \ref{lemma:minmax} are satisfied. If follows that  
\begin{align*} 
\mathcal{V}_t(x) = 
\min_{ \omega_{t+1} \in \Omega_{t+1}, ..., \omega_{N} \in \Omega_{N} }
\sup_{\tilde f_t \in \Bc_t(x) }  \frac{1}{2} x\tp Q x_t  \\
+ \tilde E [ (Ax + v_t) \tp {\rev S_{t+1}} (Ax + v_t)] + \sum_{s=t+1}^N r_{s}.
\end{align*}
Finally, exploiting Theorem \ref{th:wp_strongdual} and using the recursions \eqref{eq:wp_Riccati} and \eqref{eq_wp_z}
we have that 
$$ 
\mathcal{V}_t(x) = \min_{\omega_{t} \in \Omega_{t}, ..., \omega_{N} \in \Omega_{N}} \frac{1}{2} x\tp {\rev S_t} x + \sum_{s=t}^N r_{s}
$$
and the minimum does exists. 
Notice that ${\rev S_t} \in \Sb_{++}^n  $ in view of Lemma \ref{lemm:pd}.  
This completes our inductive argument. 
$\hfill \qed$

{\rev
The analysis conducted in this section allows us to conclude that the worst-case performance analysis problem \eqref{pb:WP} can be addressed by solving the dual problem \eqref{eq:wp_value_fcn} when $t=0.$  
This latter problem can be numerically solved by adopting a coordinate descent algorithm. 
A more detailed explanation of the coordinate descent algorithm within this framework will be provided in the following section in relation to the problem \eqref{eq:drc_value_fcn}, which shares a similar nature to \eqref{eq:wp_value_fcn}.
}

 \begin{remark}\label{rem:wc_performance}
	We immediately recognize  that our solution has the same structure of the solution to the robust performance analysis problem considered in \cite[Section IV.B]{Petersen2000}. However, instead of using a constant Lagrange multiplier $\omega,$   here $\omega_t$ is time-dependent. Indeed, whereas in the previous work \cite{Petersen2000} a single relative entropy constrained was imposed, resulting in a unique Langrange multiplier, here each $ \tilde f_t$ has an associated relative entropy constraint.  
	Moreover, by writing $\theta_t = \omega_t ^{-1} $, we easily see that the solution to our problem takes the form of a risk-sensitive cost (see the state feedback control results given in \cite{1973jacobson} in the special case where there is no control input).
	However, while in standard risk-sensitive control problems the risk sensitive parameter $\theta$ appearing in the exponential of the quadratic loss function is constant, here $\theta_t$ is time-dependent.
	As a matter of fact, in the present paper we tackle the robust analysis problem by evaluating $N+1$ risk-sensitive costs with a time-varying risk-sensitive parameter, namely
	$$
	\int \exp \big\{\frac{1}{2 \omega_{t} }  (Ax + v_t)\tp {\rev S_{t+1}}  (Ax + v_t) \big\} f_t (v_t) dx_{t} $$ 
 for $t=N, \dots, 0.$	 
As a result we obtain a risk-sensitive Riccati recursion with a time-dependent sensitivity parameter $1/\omega_t.$ 
\end{remark}


\section{Distributionally robust control policy}\label{sec:minmax}
In this section, we deal with the distributionally robust control problem \eqref{pb:DRC}.
In a similar fashion to Section \ref{sec:worst-perf-analysis}, we use the dynamic programming technique. 
Let $V_{t}: \R^n \to \R$ be the value function defined as
\begin{align*}
  V_{t} (x) := \inf_{\pi \in \Pi} \sup_{\gamma \in \Gamma} \tilde \E  \Big[  &\frac{1}{2} \sum_{s=t}^{N} (x_s\tp Q x_s + u_s\tp R u_s)   +\\
  &\frac{1}{2} x_{N+1}\tp Q_{N+1} x_{N+1} \; | x_t = x \Big].
\end{align*} 
Clearly, $V_t (x)$ represents the optimal worst-case expected cost-to-go from stage $t$ given $x_t = x.$
The value function satisfies the recursion
\begin{align*}
V_t (x)  = \inf_{u_t} \sup_{\tilde f_t \in \Bc_t(x) } \tilde \E \Big[ \frac{1}{2} (x_t \tp Q x_t + u_t \tp R u_t + \\
V_{t+1} ( Ax_t + Bu_t + v_t) | x_t = x\Big] 
\end{align*}
for $t=0, \dots, N.$
For $t = N+1 $ we have that 
$$ V_{N+1} (x) =  \frac{1}{2} x\tp Q_{N+1} x. $$
According to the dynamic programming principle,  it holds that 
$$  \inf_{\pi \in \Pi} \sup_{\gamma \in \Gamma} J(\pi,\gamma,\bar x_0)  = V_0 (\bar x_0).$$

\begin{theorem}[Optimal policy]
\label{th:drc_main}	
The value function of problem \eqref{pb:DRC} 
can be expressed as 
\beq \label{eq:drc_value_fcn} 
V_{t} (x) = \min_{\lambda_{t} \in \Lambda_t , \dots, \lambda_{N} \in \Lambda_{N} } W_{t} (x, \lambda_{t} , \dots, \lambda_{N} )  \eeq
where 
\beq  \label{eq:drc_w_function} 
W_{t} (x, \lambda_{t} , \dots, \lambda_{N}) = \frac{1}{2} x \tp P_{t}x + \sum_{s=t}^{N} c_s.
\eeq
Here, $P_t(\lambda_{t} , \dots, \lambda_{N})$ is recursively computed as 
\beq\label{eq:drc_Riccati}  
P_{t}  =  Q + \lambda_{t} E_1\tp E_1 +  A\tp (P_{t+1}^{-1} + B R\inv B\tp -  \frac{V}{\lambda_{t}} )^{-1} A 
\eeq 
for $t=N,...,0,$ with terminal condition $P_{N+1} = Q_{N+1},$
and $c_t(\lambda_{t} , \dots, \lambda_{N}) $ is defined by
\beq \label{eq:drc_z}
c_t =  -  \frac{\lambda_t}{2} \ln \left| I - \frac{P_{t+1}V}{\lambda_t} \right|  {\rev +} \lambda_t d_t.  
\eeq
Finally, 
\beq  \label{eq:drc_Lambda_set}
\Lambda_t := \big\{ 
\lambda_t \in \R_{++} \; : \; \lambda_t > \max \{\eig(P_{t+1}V) \} \big\}.
\eeq
The minimum in \eqref{eq:wp_value_fcn}   is achieved for some $(\lambda_{t}^\star, .. , \lambda_{N}^\star) \in \Lambda_t \times \dots \times \Lambda_{N}.$
Furthermore, given $P^\star_{t} = P (\lambda_{t}^\star, \dots, \lambda_{N}^\star),$ 
Problem \eqref{pb:DRC} has the unique optimal policy 
\beq \label{eq:drc_policy}
\pi^\star_{t} (x) = -K_t x,  \qquad t=0, \dots, N
\eeq 
where 
 \beq \label{eq:drc_K} 
 K_t :=  R\inv B\tp \Big( {P^\star_{t+1}}\inv + B R\inv B\tp - \frac{V}{\lambda^\star_{t} } \Big)\inv A.
 \eeq
\end{theorem}

In order to derive Theorem \ref{th:drc_main}, we state the following preliminary results.

\begin{lemma}\label{lemm:drc_pd}
    Let $P_{t+1} \in \Sb_{++}^n $ and $\lambda_t \in \Lambda_t.$ Assume that Assumption \ref{ass:obs} is satisfied.
    Then, $P_{t+1}\inv - V/\lambda_t \succ 0 $ and $P_t$ defined by \eqref{eq:drc_Riccati} is positive definite.
\end{lemma}
\proof
The proof is the same as the one of Lemma \ref{lemm:pd}, and it is therefore omitted. 
$\hfill \qed $

\begin{theorem} \label{th:drc_strongdual} 
  Given $x \in \R^n$ and $P_{t+1} \in \Sb_{++}^n $, let
\begin{align} 
\begin{aligned}
\mu  = \inf_{u} \sup_{\tilde f_t \in \Bc_t(x)} \tilde \E \big[ \frac{1}{2} &(Ax + B{\rev u} + v_t)\tp P_{t+1} \times\\  
&   (Ax + Bu + v_t)   +u\tp R u \big]  \label{pb:drc_primal} 
\end{aligned}
\end{align} 
and 
\beq
\phi   = \inf_{\lambda_t \in \Lambda_t}   \frac {1}{2} x\tp (P_t  - Q ) x\tp + c_t   \label{pb:drc_dual}
\eeq
with $P_t,$ $\Lambda_t$ and $c_t $  defined by \eqref{eq:drc_Riccati}-\eqref{eq:drc_z}.
Then $\mu = \phi $ and the minimum in \eqref{pb:drc_dual} is achieved for some $\lambda_t^\star \in \Lambda_t.$  
Moreover, the outer minimization problem in \eqref{pb:drc_primal} has a unique minimizer
$$u = - R\inv B\tp \Big( {P_{t+1}}\inv + B R\inv B\tp - \frac{V}{\lambda^\star_{t} } \Big)\inv A x. $$ 
\end{theorem}

\proof
Given $u \in \R^m, $ consider the problem
\begin{align} \label{pb:primal_partial}
\begin{aligned}
\tilde \mu (u)  = \sup_{\tilde f_t \in \Bc_t(x)}
\tilde \E [ & 
\frac{1}{2} (Ax + Bu + v_t)\tp P_{t+1}\times  
 \\ 
& (Ax + Bu + v_t) + \frac{1}{2}  u\tp R u  ]
\end{aligned}
\end{align}
Next, we derive the dual problem of \eqref{pb:primal_partial}.
Let $\lambda_t \geq 0$ be the Lagrange multiplier and    
$L(u, \tilde f_t, \lambda_t) = \tilde \E [ \frac{1}{2} (Ax + Bu + v_t)\tp P_t (Ax + Bu + v_t) ] +  \frac{1}{2} u\tp R u   - \lambda_t ( {\rev \Rc_t} - \frac{1}{2} \Vert E_1 x \Vert^2 - d_t ) 
$
be the Lagrangian of \eqref{pb:primal_partial}.
The dual function is defined as the supremum of $L (u, \tilde f_t, \lambda_t)$ over $\tilde f_t \in \Pc(\R^n).$ 
If $\lambda_t = 0, $ then $\sup_{\tilde f_t \in \Pc(\R^n) } L (u, \tilde f_t, \lambda_t ) = \infty. $
On the other hand, for $\lambda_t > 0$, we have that
\begin{align*} 
&\sup_{\tilde f_t}  L    
=  
\frac{\lambda_{t}}{2} \Vert E_1 x \Vert^2  + \lambda_{t} d_t  + \lambda_t \Big( \sup_{\tilde f_t}    \int \big( (Ax + Bu + v_t)\tp \\ 
&  \times   \frac{P_{t+1}}{2\lambda_{t}} (Ax + Bu + v_t) + 
 u\tp R u  \big )\tilde f_t (v_t) dv_t  -  {\rev \Rc( \tilde f_t || f_t)}  \Big) \\
& = \frac{\lambda_{t}}{2} \Vert E_1 x \Vert^2  + \lambda_{t} d_t  +\lambda_t  \times\\ 
&   \ln \big(  \int  e^{\frac{1}{2\lambda_{t}} \big(  u\tp R u
+  (Ax + Bu + v_t) \tp P_{t+1} (Ax + Bu + v_t)\big) } f_t (v_t) dv_t  \big)
\end{align*} 
where in the last equality we have used Lemma \ref{lemma:KLopt}.
In view of Lemma \ref{lemma:GaussIntegral}, 
the integral in the above expression converges if and only if $P_{t+1}\inv - V/\lambda_t \succ 0.$
Thus, $ \sup_{\tilde f_t} L < \infty $ only if $ \lambda_t \in \Lambda_t.$
Therefore, we can formulate the dual problem for \eqref{pb:primal_partial} as 
\beq  
\tilde \phi (u) = \inf_{\lambda_t \in \Lambda_t} D(u, \lambda_t) 
\eeq 
 where
\begin{align*} 
D& = \lambda_t  \ln \big(  \int   e^{\frac{1}{2\lambda_{t}} \big(u\tp R u +  (Ax + Bu + v_t) \tp P_{t+1} (Ax + Bu + v_t)\big) } 
 \\ 
& \times f_t (v_t) dv_t  \big)+ \frac{\lambda_{t}}{2} \Vert E_1 x \Vert^2  + \lambda_{t} d_t. 
\end{align*}
Similarly to the proof of Theorem \ref{th:wp_strongdual}, we can use Theorem \ref{lemma:luenberger} to conclude that
$\tilde \mu (u) = \tilde \phi (u).$  
Hence, 
$$ 
\mu = \inf_{u} \tilde \mu (u) 
 = \inf_{\lambda_t \in \Lambda_t} \inf_{u} D(u, \lambda_t).$$ 
Notice that, fixed $\lambda_t \in \Lambda_t$, the problem
$ \inf_{u}  D(u, \lambda_t)  $
is a one-step LQG risk-sensitive optimal control problem.
The unique optimal control strategy for this problem  is obtained by Jacobson in \cite{1973jacobson} and it is 
$ u(\lambda_{t}) = - R\inv B\tp ( {P_{t+1}}\inv + B R\inv B\tp - V / \lambda_{t})\inv A x$.
Correspondingly, 
$ \tilde D(\lambda_t) := \inf_{u} D(u, \lambda_t) = x\tp A\tp (P_{t+1}^{-1} + B R\inv B\tp -  \frac{V}{\lambda_{t}} )^{-1} A x -  \frac{\lambda_{t}}{2} \ln | I - \frac{P_{t+1}V}{\lambda_t} | +  \frac{\lambda_{t}}{2} \Vert E_1 x \Vert^2  + \lambda_{t} d_t. $
Exploiting this result, we obtain that 
$$
\mu = \inf_{\lambda_t \in \Lambda_t} \tilde D(\lambda_t).  
$$
To complete the proof, it remains to show that Problem \eqref{pb:drc_dual} admits a minimum point.
To this end, we show that we can restrict the search of the infimum over a compact subset of $\Lambda_t$. Consider a sequence $(\lambda_t^{(k)})_{k \in \N} \in \Lambda_t $ such that $\lambda_t^{(k)} \to \infty$ as $ k \to \infty.$ 
This implies that $\tilde D(\lambda_t^{(k)}) \to \infty$ as $ k \to \infty.$ Hence, this sequence cannot be an infimizing sequence.
Furthermore, consider a sequence $(\lambda_t^{(k)})_{k \in \N} \in \Lambda_t $ such that $\lambda_t^{(k)} \to \max(\eig(P_{t+1}V))$ as $ k \to \infty.$
This implies that, as $ k \to \infty,$ $|I - P_{t+1}V/\lambda_t^{(k)}| $ tends to zero, and hence $ - \frac{\lambda_t^{(k)}}{2} \ln | I - \frac{P_{t+1}V}{\lambda_t^{(k)}} | \to \infty.$ Therefore, $\bar D(\lambda_t^{(k)})$ tends to infinity and the considered sequence cannot be an infimizing sequence.
Thus, the minimization over $\Lambda_t$ is equivalent to the minimization over the subset $ \{ \lambda_t \in {\rev \R_{++}} : \alpha \leq \lambda_t \leq \beta \} $ with $\alpha, \beta \in \R_{++},$ $\alpha > \max(\eig(P_{t+1}V)).$ 
Then, since the objective function is continuous over the restricted compact set, in view of  Weierstrass's theorem, we conclude that problem does admit a minimum.

$\hfill \qed$

\begin{lemma}\label{lemma:drc:conv}
Given $x \in \R^n,$ the function $W_t (x, \lambda_{t}, .., \lambda_N) $ defined by \eqref{eq:w_function} is jointly convex in $(\lambda_{t}, .., \lambda_{N}) \in \Lambda_t \times .. \times \Lambda_N$ for any $t=0,\dots, N.$
\end{lemma}
\proof  
By recursively applying Theorem \ref{lemma:KLopt} and the solution to the risk-sensitive LQG control problem, it is possible to verify that the function $W_t$ can be equivalently written as
\begin{align*}
W_t = \inf_{u_t, \dots, u_N} \sup_{ \tilde f_t, \dots , \tilde f_N \in  \Pc(\R^n)}    \tilde \E 	 \frac{1}{2}	\Big[  \sum_{k=t}^{N}  x_k\tp Q x_k  + u_k \tp R u_k +  \\ 
x_{N+1}\tp Q_{N+1} x_{N+1} | x_t = x\Big] 
- \sum_{k=t}^{N} \lambda_k ({\rev \Rc_k} -  \frac{1}{2} \Vert z_k \Vert ^ 2  - d_k)
\end{align*}
for $(\lambda_{t}, .., \lambda_{N}) \in \Lambda_t \times .. \Lambda_N$ .   
In the latter minimax optimization problem, the objective function is jointly convex in $(u_t, .., u_N, \lambda_{t}, .., \lambda_{N})$ for any $(\tilde f_t , .., \tilde f_N ) $. Then, by Lemma \ref{lemma:convex}
the maximization with respect to $(\tilde f_t, .., \tilde f_N) $ returns a convex function of 
$(u_t, .., u_N, \lambda_{t}, .., \lambda_{N}).$ Finally, in view of Lemma \ref{lemma:convex_min}, the partial minimization over the control input variables $(u_t, .., u_N)$ 
preserves convexity with respect to $(\lambda_{t}, .., \lambda_{N}).$ 
$\hfill \qed$

\emph{Proof of Theorem \ref{th:drc_main}}
The theorem can be proved via mathematical induction.
The induction hypothesis holds at $N+1$. 
Given the induction hypothesis at $t+1,$ after using Sion's minimax Theorem \ref{lemma:minmax}, we obtain  
\begin{align*} 
V_t(x) = \inf_{u} \min_{\lambda_{t+1} \in \Lambda_{t+1}, \dots \lambda_{N} \in \Lambda_{N+1} }  \sup_{\tilde f_t \in \Bc (x) }  \frac{1}{2} x\tp Q x \\
+ u\tp R u + 
\tilde E [W(Ax + Bu + v_t, \lambda_{t+1}, \dots \lambda_{N} )]
\end{align*}
Now, to proceed in the mathematical induction, we need to switch the minimization with respect to $u_t $ and the minimization with respect $\lambda_{t+1}, .. \lambda_N$, obtaining 
\begin{align*}
V_t(x) = \inf_{\lambda_{t+1} \in \Lambda_{t+1}, \dots \lambda_{N} \in \Lambda_{N}}
\inf_{u}   \sup_{\tilde f_t \in \Bc (x) }  \frac{1}{2} x\tp Q x + u\tp R u + \\
\tilde E [W(Ax + Bu + v_t, \lambda_{t+1}, \dots \lambda_{N}) ]
\end{align*}
Finally, from Theorem \ref{th:drc_strongdual} it follows that 
$$
V_t(x) = \inf_{\lambda_t \in \Lambda_t, \lambda_{t+1} \in \Lambda_{t+1}, \dots \lambda_{N} \in \Lambda_{N}}
W_t(x, \lambda_t, ...,\lambda_N).
$$
The fact that the infimum in the last formula is actually a minimum can be proved by using similar arguments to the last part of the proof of Theorem \ref{th:drc_strongdual}. 
$\hfill \qed$

From Theorem \ref{th:drc_main} we conclude that
$$ \inf_{\pi \in \Pi} \sup_{\gamma \in \Gamma} J(\pi,\gamma, \bar x_0)  = \min_{\lambda_{0} \in \Lambda_0 , \dots, \lambda_{N} \in \Lambda_{N} } W_0 (\bar x_0, \lambda_{0}, \dots, \lambda_{N}). $$
{\rev
We propose a coordinate descend algorithm \cite{Spall2021} to solve the problem on the right hand side of the previous equation. 
Since this problem is convex, the coordinate descent algorithm converges to the optimal point. 
At each iteration, we need to solve a single-variable optimization problem in $\lambda_t$. Due to its convexity, methods such as bisection or grid search can be effectively employed.
To do that, we solve the Riccati equations \eqref{eq:drc_Riccati} for different values of $\lambda_t,$ then we evaluate the quantity $ W_0 (\bar x_0, \lambda_{0}, \dots, \lambda_t, \dots, \lambda_{N})$ as a function of $\lambda_t$ and we choose the value of the variable for which the minimum is achieved. 
The overall procedure to compute the D$^2$-LQG controller exploiting the coordinate descend method is summarized in Algorithm \ref{algo:coord_desc}.
\\
From Algorithm \ref{algo:coord_desc} and Theorem \ref{th:drc_main}, it should be  clear that computing the optimal control gain $K_t$ requires the knowledge of the matrix $P^\star_{t+1}$  from the next time step.
This is not an issue, as $P^\star_{t+1}$ has been already computed as part of the process. 
}

\begin{algorithm*}
\caption{D$^2$-LQG controller}\label{algo:coord_desc}
\begin{algorithmic}
{\rev
\Require $A,B,Q,Q_{N+1},R,V,E_1,\{d_t\}_{t=0}^{N}$
\State Initialize $\lambda^0 = (\lambda_{0}^0, \dots, \lambda_{N}^0) \in \Lambda_0 \times \cdots \Lambda_N$
\State Set $k=0$
\Repeat \Comment{\textit{coordinate descend iterations}}
    \For{$j=0$ to $N$} 
        \State 
        $\lambda_j^{k+1} = \argmin_{y \in \Lambda_j} W_0(\bar{x}_0, \lambda_0^{k+1}, \dots, \lambda_{j-1}^{k+1}, y , \lambda_{j+1}^{k}, \dots,  \lambda_N^{k}) $
        \Comment{\textit{use bisection method}}
    \EndFor
    \State $k = k+1$
\Until{convergence} 
\State Set $(\lambda_{0}^\star, \dots, \lambda_{N}^\star) = (\lambda_{0}^k, \dots, \lambda_{N}^k)$
\State Set $P_{N+1}^\star = Q_{N+1}$
 \For{$t=N$ to $0$} 
        \State Compute $P_t^\star$ using $\lambda^\star_t$ and $P_{t+1}^\star$ via \eqref{eq:drc_Riccati}
        \State Compute $K_t$ using $\lambda^\star_t$ and $P_{t+1}^\star$ via \eqref{eq:drc_K}
    \EndFor
}
\end{algorithmic}
\end{algorithm*}

A salient feature of the solution scheme presented in Theorem \ref{th:drc_main} is that an explicit form of the maximizers players $\tilde f_t^\star(v_t)$ in Problem \eqref{pb:DRC} is not required to obtain the optimal control law.
However, for simulation and performance evaluation, it is instructive to construct the least favorable model. 

\begin{corr}[Worst-case distribution]
The policy $\gamma^\star \in \Gamma $ achieving the maximum in Problem \eqref{pb:DRC} is 
$ \gamma^\star_t (x,u) = \tilde f_t^\star$ for $t=0, \dots, N,$ with 
$
\tilde f_t^\star   =   
\Nc \Big( 
\big( 
V\inv  - 
\frac{P_{t+1}^\star}{\lambda_t^\star} 
\big)^{-1}
\frac{P_{t+1}^\star}{\lambda_t^\star} (Ax + Bu)  
,  
\big( V\inv - 
\frac{P_{t+1}^\star}{\lambda_t^\star} \big)^{-1}
\Big). 
$
\end{corr}
\proof
The result is a consequence of Theorem \ref{th:drc_main} and Theorem \ref{th:drc_strongdual}, by noticing that at each step of the dynamic programming recursion, we leveraged on Lemma \ref{lemma:KLopt} to evaluate the worst-case scenario (see the proof of Theorem \ref{th:drc_strongdual}). 
This lemma implies that, given $\lambda_t \in \Lambda_t,$ $x \in \R^n$ and $ u \in \R^m,$ the admissible worst-case  distribution is given by
$$
\tilde f_t^\star =   
\frac{  
e^ {\frac{1}{2\lambda_t } ( Ax + Bu + v_t ) P_{t+1} ( Ax + Bu + v_t ) } f_t 
}
{ \int e^ {\frac{1}{2\lambda_t } ( Ax + Bu + v_t ) P_{t+1}^\star (Ax + Bu  + v_t ) } f_t(v_t) dv_t } .
$$
The thesis follows by applying Lemma \ref{lemm:worstpdf} and substituting $\lambda_t = \lambda_t^\star$ and $P_{t+1} = P_{t+1}^\star $ 
$\hfill \qed$

It is clear that the least-favorable density of the noise $v_t$ involves a perturbation of both the mean and the variance of the nominal noise distribution.
Notice also that for the worst-case distribution the noise random variables $v_t$ and $v_s$ with $t \neq s$ are not independent.

\begin{remark}
We have assumed in Section \ref{sec:pb} that the admissible control policies $\pi_t$ and the adversary policies $\gamma_t $  are functions of the state vector only at the current time step $t.$ This hypothesis is not restrictive. 
In fact, we can repeat the previous argument under the assumption that the controller and the adversary have access to the whole past history: the computations are essentially the same and the  optimal control policy does not change. 	In other words, no additional information about the future development of the system is obtained if past measurements are included. This is indeed rather intuitive if we consider the Markovian property of the state.
\end{remark}	

{\rev 
\begin{remark}
    Similarly to Remark \ref{rem:wc_performance}, we observe that our solution
    share the same structure as the solution to the DRC control problem considered in \cite{Petersen2000} and the risk-sensitive LQG controller \cite{1973jacobson}. 
    Assuming for simplicity that $E_1 = 0,$ we also note that the Riccati recursion \eqref{eq:drc_Riccati} and the control gain \eqref{eq:drc_K} are similar to those in the standard LQG problem, 
    with the key difference being the presence of the term $\frac{V}{\lambda_t^\star}.$
    This term not only accounts for the nominal noise distribution $V$, but also includes information about the ambiguity set and the worst-case model, as captured by the optimal Lagrange multiplier $\lambda_t^\star.$  
    The term $\frac{V}{\lambda_t^\star}$ effectively controls the trade-off between optimality and robustness. When the radius of $\mathcal{B}_t$ approaches zero,  $\lambda_t^\star$ tends to infinity, making $\frac{V}{\lambda_t^\star}$ negligible. As a result, the D$^2$-LQG controller converges to the standard LQG controller.
    On the other hand, as the radius of $\mathcal{B}_t$ increases, the difference between the D$^2$-LQG controller and the standard LQG controller becomes more significant. In the extreme situation in which the radius of  $\mathcal{B}_t$ grows to infinity, the D$^2$-LQG controller becomes highly conservative and pessimistic.
\end{remark}
}

{\rev
\begin{remark}
If the ambiguity set $\Bc_t $ includes the true underlying model at each time step $t$, the proposed D$^2$-LQG control \eqref{eq:drc_policy} ensures that the expected closed-loop cost is smaller than $W_0(\bar x_0, \lambda_0^\star, \dots, \lambda_N^\star).$
However, in practical scenarios, the ambiguity set is often constructed from some a priori knowledge or an estimate of the uncertainty in the system. Thus, it is possible that the real underlying system does not belong to this set. 
In these situations, the control law \eqref{eq:drc_policy} remains applicable, but the upper-bound performance guarantees are lost.
\end{remark}}

\section{Simulations} \label{sec:simulation}
We consider the control of an inverted pendulum on a cart, which is a typical benchmark in control theory.
The inverted pendulum system is depicted in Figure \ref{fig:pendulum}. It consists of a pendulum which is attached to a cart equipped with a motor driving it.
The movement of the cart is constrained to the horizontal direction, whereas the pendulum can rotate in the vertical plane. 
The motion of the system can be represented by the following equations \cite{pendulum}
\beq \begin{aligned} \label{eq:invertedpend_nl}
m_pl \ddot{\theta} \cos \theta - m_p l \dot{\theta}^2  \sin \theta + (M_c+m_p) \ddot{q} = u \\
J \ddot{\theta}  - m_p lg \sin \theta + m_p l \cos \theta \ddot{q} = 0,
\end{aligned}\eeq
where $\theta$  (in \si{\radian}) is the angle between the vertical axes and the rod of the pendulum,  $q$ (in \si{\metre}) is the horizontal displacement of the pendulum  and $u$ (in \si{\newton}) is the input acceleration; $M_c = 3  \; \si{\kilogram} $ is the mass of the cart, $m_p = 1.5 \; \si{\kilogram} $ is the mass of the pendulum, $l = 2 \; \si{\metre}$
is half of pendulum length, $g = 9.8  \; \si{\metre / \second }$
is the gravity acceleration, and
 $J= \frac{4}{3} ml^2$  is the pendulum moment of inertia around the pivot. 
The objective is to design a control law stabilizing the pendulum around the unstable equilibrium point $ \theta = 0$ while the cart executes a uniform linear motion with velocity $\dot{q}_d.$
Let 
$x = [\theta \; \dot{\theta } \; q \; \dot{q} ] \tp $
be the state variables and 
$x_d = [0 \; 0 \; q_d \; \dot{q}_d ]\tp $ be the desired trajectory. 
By linearizing model \eqref{eq:invertedpend_nl} around the desired trajectory, we obtained the  linear state-space model \cite{pendulum}
\beq \label{eq:invertedpend_lin} 
\dot{\bar x} = \m{ 0 & 1 & 0 & 0 \\ \frac{3(M_c+m_p)g}{4M_c l + m_p l}	& 0 & 0 & 0\\ 
	0 & 0 & 0 & 1 \\ -\frac{3m_p g}{4M_c +m_p} & 0 & 0 & 0} \bar x + \m{0 \\ -\frac{3}{4 M_c l + m_p l} \\ 0 \\\frac{4}{4M_c + m_p} } u ,
\eeq
where $\bar x := x - x_d $ is the tracking error.
\\
We are warned that the cart has to cross an irregular ground made of cobblestones in the time interval $t \in [15,45]$
(given that the cart is moving with constant velocity $\dot q_d).$ 
Moreover, we know that model \eqref{eq:invertedpend_lin} is an approximate representation of the system dynamics, because for example it does not take into account non-linearities and friction effects. 
Hence, the need to design a control policy which is robust against uncertainties and  mis-specifications.
To this end, we first discretize model \eqref{eq:invertedpend_lin}  using zero-order hold on the input and a sample time of $T_s = 0.2 \sec.$ 
The resulting state and input matrices are
\beq \label{eq:pendulum_AB}
A = \m{ 1.099 & 0.206 & 0 & 0\\
	1.012   & 1.099 & 0 & 0 \\
	-0.066  &  -0.004  & 1 & 0.2 \\
	-0.675   &-0.066 & 0 & 1},
\;  B =  \m{-0.0023 \\	-0.0230 \\	0.0060 \\ 0.0597}.
\eeq
Then, we apply the proposed D$^2$-LQG procedure to control the system. 
Specifically, we consider a time interval of length $N = 100$ 
with initial condition
$\bar x_0 = [0.1, \; \; -0.1, \; \;0.05, \; \; 0.02]\tp.$
The quadratic cost is specified by the matrices $Q = Q_{N+1} = \diag(  10, 1, 10, 1 ) $ and $R = 1.$
We define the ambiguity set \eqref{eq:amb_set} by setting $ f_t = \Nc (0, V)$ with $V=\diag(0.1,0.5,0.1,0.5)$ and $E_1 =[0.5 \; \; 0 \;\; 0.5 \; \;0].$
Finally, to take into account the cobblestones ground, we set $d_t  =  0.2 $ for  $t \in [15,45] $ and $d_t = 0.001 $ otherwise. 
{\rev Applying Algorithm 1, the computation of the D$^2$-LQG control policy required 60.83 \si{\second } on a processor Intel\textregistered \hspace{0.01cm} 
 Core\texttrademark  \hspace{0.01cm}  i7-1355U. } 
For the sake of comparison, we also compute the control policy obtained by the standard LQG scheme
as well as the distributionally robust control scheme proposed in \cite{Petersen2000} where the ambiguity set is defined as 
$
\Bc = \{ \tilde f :  
{\rev \Rc( \tilde f || f  ))} \leq d +  \tilde \E  [ \; \sum_{t=0}^{N} \frac{1}{2}\Vert z_t \Vert^2 ] \},
$
where we recall that $\tilde f $ and $ f $ are the perturbed and the nominal probability distributions of the noise sequence $ v = [v_0, ..., v_N]$.  
Notice that, for a fair comparison, we set 
$ d = \sum_{t=0}^{N} d_t.$
The latter control scheme will be hereafter denoted as D-LQG.
\\
To assess the performance of the three control schemes in terms of optimality and robustness, we assume that the real dynamics of the inverted pendulum system is 
\beq \label{eq:invertedpend_lin_DT_per}
x_{t+1} = (A+ \Delta A) x_t + B u_t + v_t,
\eeq
with $\Delta A$ representing the unmodeled dynamics and $v_t$ modeling the effect of the irregular ground. 
We assume that $v_t \sim \Uc( 0, \diag( 0,   0.0125 , 0 , 0.125))  $ for $t = [15, 45]$ and $v_t = 0$ otherwise.
We perform three Monte Carlo experiments corresponding to three different choices of $\Delta A.$ Each experiment is composed by $1000$  trials and in each trial we draw a realization of $v_t$. 
{ \rev The results are summarized in Table \ref{tab:pendulum_results}, which displays the average closed-loop cost achieved by the three methods across the Monte Carlo simulations.  
The results reveal that the system controlled by the standard LQG technique fails to stabilize the system, as evidenced by the dramatically large mean cost values, which exceeds $10^6$ for all three systems. 
Additionally, the proposed distributed approach outperforms  Petersen's single constraint approach, reaching a  lower average cost for all three systems. 
 The table also shows the highest ratio $r_{max} $ and the lowest ratio $r_{min} $ between the $D$-LQG and D$^2$-LQG cost over the Monte Carlo experiments.   Notably,  $r_{min} $  and $r_{max}$ are greater than one for all the simulations. This indicates that the D$^2$-LQG policy consistently outperforms the $D$-LQG policy, not only on average, but in each individual Monte Carlo experiment. 
}
\\
It is also interesting to compare the performance  of the three control schemes when the system to be controlled is nominal one. 
In other words, we assume that the inverted pendulum evolves according to the dynamics
\beq \label{eq:pendulum_nom}
x_{t+1} = Ax_t + Bu_t
\eeq
with $A$ and $B$ defined in \eqref{eq:pendulum_AB} and we apply the LQG, the D-LQG controller and the D$^2$-LQG controller.
The results are summarized in Table \ref{tab:pendulum_nominal}.
{\rev Under nominal conditions, we know from the theory that the LQG control policy is the optimal one. As a matter of fact, it reaches the lowest value of the closed-loop cost. Note that the proposed D$^2$-LQG approach reaches an average closed-loop cost that lies between the standard LQG and the D-LQG controllers.  Specifically, while it does not match the minimal cost achieved by the LQG controller, it performs better than the D-LQG controller.
\\
The results in Tables \ref{tab:pendulum_results} and \ref{tab:pendulum_nominal} confirm that the proposed D$^2$-LQG approach effectively balances optimality and robustness. Indeed, it mitigates the conservatism inherent in the D-LQG approach, achieving a lower cost across all simulations. At the same time,  it ensures robustness, maintaining good performance even when there is a mismatch between the real underlying system and the nominal one.
}
\\
Notice that we have repeated the simulations  for different values of $E_1$ and $d_t$, obtaining similar results to the previous ones. 
\\
{\rev
Finally, we remark that the previous example stretches
our theory beyond the limits of our mathematical assumptions, as
we have not made any specific assumptions on the perturbation  matrix $\Delta A$ in \eqref{eq:invertedpend_lin_DT_per}. 
This means that we do not have guarantees that the underlying true model lies within the ambiguity set that  we have defined. In fact, a numerical analysis reveals that the relative entropy constraint is violated for some time steps across all three choices of $\Delta A$. 
Our simulations show that the controller remains robust despite this mismatch.
In conclusion, our controller may be employed in practical scenarios
where the ambiguity set is given by deterministic parameters variations and hence cannot be easily encapsulated into a ball in the relative entropy topology.
}


\begin{figure} 
	\centering
	\tikzset{every picture/.style={line width=0.75pt}} 
	\begin{tikzpicture}[x=0.75pt,y=0.75pt,yscale=-0.7,xscale=0.7]
		\draw   (264,173) -- (354,173) -- (354,213) -- (264,213) -- cycle ;
		\draw   (320.8,226.6) .. controls (320.8,219.64) and (326.44,214) .. (333.4,214) .. controls (340.36,214) and (346,219.64) .. (346,226.6) .. controls (346,233.56) and (340.36,239.2) .. (333.4,239.2) .. controls (326.44,239.2) and (320.8,233.56) .. (320.8,226.6) -- cycle ;
		\draw   (302.5,172.73) .. controls (302.18,171.94) and (302,171.09) .. (302,170.2) .. controls (302,166.33) and (305.33,163.2) .. (309.44,163.2) .. controls (313.55,163.19) and (316.88,166.33) .. (316.89,170.19) .. controls (316.89,171.08) and (316.71,171.93) .. (316.39,172.72) -- cycle ;
		\draw  [color={rgb, 255:red, 0; green, 0; blue, 0 }  ,draw opacity=1 ] (361.81,47.83) -- (367,50.2) -- (314.81,164.51) -- (309.62,162.14) -- cycle ;
		\draw [color={rgb, 255:red, 128; green, 128; blue, 128 }  ,draw opacity=1 ] [dash pattern={on 4.5pt off 4.5pt}]  (310,28.2) -- (310,260.2) ;
		\draw   (271.8,226.6) .. controls (271.8,219.64) and (277.44,214) .. (284.4,214) .. controls (291.36,214) and (297,219.64) .. (297,226.6) .. controls (297,233.56) and (291.36,239.2) .. (284.4,239.2) .. controls (277.44,239.2) and (271.8,233.56) .. (271.8,226.6) -- cycle ;
		\draw [color={rgb, 255:red, 128; green, 128; blue, 128 }  ,draw opacity=1 ]   (310,55.2) .. controls (323,48) and (338.33,48) .. (353,58) ;
		\draw [shift={(353,58)}, rotate = 209.05] [color={rgb, 255:red, 128; green, 128; blue, 128 }  ,draw opacity=1 ][line width=0.75]    (10.93,-3.29) .. controls (6.95,-1.4) and (3.31,-0.3) .. (0,0) .. controls (3.31,0.3) and (6.95,1.4) .. (10.93,3.29)   ;
		\draw    (202,240) -- (411,240) ;
		\draw [color={rgb, 255:red, 128; green, 128; blue, 128 }  ,draw opacity=1 ]   (310,249.2) -- (345,249.2) ;
		\draw [shift={(347,249.2)}, rotate = 180] [color={rgb, 255:red, 128; green, 128; blue, 128 }  ,draw opacity=1 ][line width=0.75]    (10.93,-3.29) .. controls (6.95,-1.4) and (3.31,-0.3) .. (0,0) .. controls (3.31,0.3) and (6.95,1.4) .. (10.93,3.29)   ;
		\draw [color={rgb, 255:red, 128; green, 128; blue, 128 }  ,draw opacity=1 ]   (327,167.2) -- (378,54.2) ;
		\draw [color={rgb, 255:red, 128; green, 128; blue, 128 }  ,draw opacity=1 ]   (386,58.2) -- (371,51.2) ;
		\draw [color={rgb, 255:red, 128; green, 128; blue, 128 }  ,draw opacity=1 ]   (335,171.2) -- (320,164.2) ;
		\draw  [color={rgb, 255:red, 0; green, 0; blue, 0 }  ,draw opacity=1 ][fill={rgb, 255:red, 0; green, 0; blue, 0 }  ,fill opacity=1 ] (359,40.7) .. controls (359,35.45) and (363.25,31.2) .. (368.5,31.2) .. controls (373.75,31.2) and (378,35.45) .. (378,40.7) .. controls (378,45.95) and (373.75,50.2) .. (368.5,50.2) .. controls (363.25,50.2) and (359,45.95) .. (359,40.7) -- cycle ;
		\draw [color={rgb, 255:red, 128; green, 128; blue, 128 }  ,draw opacity=1 ]   (222,190) -- (260,190) ;
		\draw [shift={(260,190)}, rotate = 179.73] [color={rgb, 255:red, 128; green, 128; blue, 128 }  ,draw opacity=1 ][line width=0.75]    (10.93,-3.29) .. controls (6.95,-1.4) and (3.31,-0.3) .. (0,0) .. controls (3.31,0.3) and (6.95,1.4) .. (10.93,3.29)   ;
		
		\draw (324,250.6) node [anchor=north west][inner sep=0.75pt]  [color={rgb, 255:red, 128; green, 128; blue, 128 }  ,opacity=1 ]  {$q$};
		\draw (327,30.4) node [anchor=north west][inner sep=0.75pt]  [font=\normalsize]  {$\textcolor[rgb]{0.5,0.5,0.5}{\theta }$};
		\draw (355,113.6) node [anchor=north west][inner sep=0.75pt]    {$\textcolor[rgb]{0.5,0.5,0.5}{2l}$};
		\draw (310,183.4) node [anchor=north west][inner sep=0.75pt]  [color={rgb, 255:red, 128; green, 128; blue, 128 }  ,opacity=1 ]  {$M_{c}$};
		\draw (377,17.4) node [anchor=north west][inner sep=0.75pt]    {$\textcolor[rgb]{0.5,0.5,0.5}{m_{p}}$};
		\draw (234,169.4) node [anchor=north west][inner sep=0.75pt]    {$\textcolor[rgb]{0.5,0.5,0.5}{u}$};	
	\end{tikzpicture}
	\caption{Inverted pendulum system.} 
	\label{fig:pendulum}
\end{figure}


\begin{table*}
\caption{Control of the inverted pendulum on a cart: comparison of the closed-loop cost when the underlying system is governed by the perturbed  dynamics \eqref{eq:invertedpend_lin_DT_per} and it is controlled by the standard LQG controller, the single-constraint distributionally robust controller (D-LQG) \cite{Petersen2000} and the proposed D$^2$-LQG controller.
The table shows the average closed-loop cost achieved by the three control schemes  among 1000 Monte Carlo experiments. The indexes $r_{max}$ and $r_{min}$  represents the greatest and the smallest ratio  between the $D$-LQG and D$^2$-LQG cost over the Monte Carlo experiments.
}\label{tab:pendulum_results}
\centering
\vskip 0.2cm
\begin{tblr}{|l|c|c|c|c|c|}		
  \hline 
  {} & \SetCell[c=3]{c}{ \textbf{Mean  cost}}  & & & $\mathbf{r_{max}}$ & $\mathbf{r_{min}}$  \\	
  {}  & LQG & $D$-LQG & D$^2$-LQG & {}  &  {}  \\	 
  \hline
    $ 
    \Delta A = 
    \m{       
    0.0269  &  0.0316 &  -0.0243  &  0.0288 \\
    -0.0296 &  -0.0290 &  -0.0163 &  -0.0312\\
    0.0332  &  0.0046 &   0.0348  & -0.0093 \\
    -0.0246  & -0.0042  & -0.0284  &  0.0138 } 
    $ 
    &  
    $1.01 \cdot 10^6 $ & $ 1.42 \cdot 10^3  $ 
    &
    $1.02 \cdot 10^3  $&  $1.54 $ &  $1.23 $ \\ 
\hline
$\Delta A =
\m{ 0.0102   & 0.0125 &   0.0231   & 0.0135 \\
    0.0160 &   0.0116 &  -0.0064   & 0.0132 \\
   -0.0162&   -0.0161 &  -0.0086  &  0.0083 \\
   -0.0006 &  -0.0018  &  0.0051   &  0.0021} $ 
    &  $ 3.80 \cdot 10^{11}    $ & $1.04 \cdot 10^3 $ &
    $ 7.85 \cdot 10^2  $ &  $1.57 $ &  $1.29 $ \\ 
 \hline 
 $\Delta A =
\m{ -0.0002 & -0.0175  &  0.0085 &  0.0029\\
    0.0081  &  0.0153   &  0.0112  &  0.0086\\
    0.0091  & -0.0271    & -0.0164   & -0.0137\\
    0.0063  & -0.0104   &  0.0099  & -0.0011 } $ 
    &  $ 5.05 \cdot 10^6  $ & $ 9.44 \cdot 10^2 $ &
    $ 6.46 \cdot 10^2  $ &  $1.59 $ &  $1.36 $ \\
    \hline
\end{tblr}
\end{table*}

\begin{table}

\caption{Control of the inverted pendulum on a cart: comparison of the closed-loop cost when the underlying system is governed by the nominal  dynamics \eqref{eq:pendulum_nom} and it is controlled by the standard LQG controller, the single-constraint distributionally robust controller (D-LQG) and the proposed D$^2$-LQG controller.
The table shows the closed-loop cost achieved by the three control schemes.
}\label{tab:pendulum_nominal}
\vskip 0.2cm
\centering
\begin{tblr}{|c| c| }
			\hline
			\rule{0pt}{2.5ex}
			\textbf{Control Scheme} & \textbf{$J(\bar x_0)$} \\ 
			\hline
			\rule{0pt}{2.5ex}
			LQG & 40.64 \\ 
			D-LQG & 729.41	\\ 
			D$^2$-LQG & 384.04 \\
			\hline
\end{tblr}		
\end{table}

\section{Extension}\label{sec:e2}
In Section \ref{sec:pb} we have assume that the tolerance of ambiguity set is given by a mismatch budget $d_t$,  independent of $x$, plus a mismatch budget proportional to the norm of the state. 
To add more flexibility in modeling system uncertainty, a natural extension consists in defining the ambiguity set as 
\beq  \label{eq:amb_set_e2}
\Bc_t (x_t, u_t) \! = \!  \big \{ \tilde f_t \in  \Pc (\R^n) : {\rev \Rc( \tilde f_t || f_t)} \leq d_t \! + \! \frac{\Vert z_t \Vert^2}{2} \!  \big\} 
\eeq
where now 
\beq \label{eq:zt_e2}  z_t = E_1 x_t + E_2 u_t, \eeq
with $ E_1 \in \R^{p \times n}$ and $E_2 \in \R^{p \times m}.$
To simplify the computations, we assume that $ E_1\tp E_2 = 0. $
Consider the D$^2$-LQG problem where $ \gamma_t$  maps the pair $(x_t, u_t)$ to the set  $\Bc_t (x_t, u_t)$ defined in \eqref{eq:amb_set_e2}.
In this scenario, the following weaker version of Theorem \ref{th:drc_main} holds. 
\begin{theorem}
The value function $V_{t} (x) $ of Problem \eqref{pb:DRC} 
is such that 
\beq 
V_{t} (x) \leq \min_{\lambda_{t} \in  \Lambda_t , \dots, \lambda_{N} \in \Lambda_{N} } W_{t} (x, \lambda_{t} , \dots, \lambda_{N} )  \eeq
where $W_{t} $ is given by \eqref{eq:drc_w_function} 
with   
$ P_t(\lambda_{t} , \dots, \lambda_{N})$ recursively computed as 
$$
 P_{t} \!  = \!   Q + \lambda_{t} E_1\tp \!  E_1 +  A\tp \! 
 (P_{t+1}^{-1} \!  + \!  B (R + \lambda_t E_2\tp \! E_2) \inv \!  B\tp \! -  \frac{V}{\lambda_{t}} 
 )^{\! -1} \!  A 
$$
for $t=N,...,0$ and terminal condition $ P_{N+1} = Q_{N+1}.$
Therefore, 
\beq \label{pb:E2_upperbound}
\inf_{\pi \in \Pi} \sup_{\gamma \in \Gamma} J(\pi,\gamma,\bar x_0) \leq \min_{\lambda_{0} \in  \Lambda_0 , .., \lambda_{N} \in  \Lambda_{N} } W_{0} (\bar x_0,  \lambda_{0} , .., \lambda_{N} ).
\eeq
Let $(\lambda_{0}^\star, .. , \lambda_{N}^\star) \in  \Lambda_0 \times \dots \times \Lambda_{N}$ achieve the minimum in the right hand-side of \eqref{pb:E2_upperbound}
and let $ P^\star_{t} = P (\lambda_{t}^\star, \dots, \lambda_{N}^\star).$ Then, the control policy 
\beq \label{eq:policy_e2}
\pi^\star_{t} (x) = - K_t x, 
\eeq 
with 
$ K_t =  R\inv B\tp ( {P^\star_{t+1}}\inv + B (R + \lambda_t^\star E_2\tp E_2) \inv B\tp - \frac{V}{\lambda^\star_{t} } )\inv A,$  is such that the corresponding cost equals the  upper bound in the right-hand side of \eqref{pb:E2_upperbound}.
\end{theorem}
\proof
The proof follows similar steps to the proof of Theorem \ref{th:drc_main}. 
The main difference is that the function $ W_t(x, \lambda_{t+1}, .., \lambda_{N+1}) $ may be non-convex in $ (\lambda_{t+1}, ..., \lambda_N) $ for $E_2 \neq 0.$
Indeed, it is not possible to extend the proof of Lemma \ref{lemma:drc:conv} when $z_t$ is given by \eqref{eq:zt_e2} with $E_2 \neq 0.$ 
As a consequence, the switch between the minimization with respect to $ (\lambda_{t+1},..., \lambda_N) $ and the maximization with respect to $\tilde f_t \in  \Bc_t $ may generate an inequality sign `` $\leq$ '' (see \cite[p.238]{boyd:vandenberghe:2004}).

\subsection{Simulations}
In the general case considered in this section, only  suboptimality of the solution can, in general, be proven.
In extensive simulations, however,  the method appears to provide 
very good performance with respect to existing alternatives.
Next we illustrate a numerical example highlighting this fact.
Consider the linear model \eqref{eq:refsys}  
with 
\begin{equation*}
A = \m{ 0.5773 &  -0.6335  & -0.0457 \\
	0.5477  &  1.7583 &   0.0524 \\
	-0.4011 &  -0.4754   & 1.0043 }, \quad    
B =   \m{0.3212 \\     0.3689 \\    -0.2741}\\
\end{equation*}
and  nominal noise variance $V = I.$
Let $ \bar x_0 = [0.5\; \; 	0.1 \;  -0.7]\tp$ be the initial state of the system.
We consider a time interval of length $N+1 = 100$ and  we measure the performance in terms of the cost \eqref{eq:cost} where $Q = 0.0025\cdot \mathbb{1}_3$, with $\mathbb{1}_3$ being a square matrix of dimension 3 with all entries equal to 1,  $Q_{N+1} = 0.1 \cdot I, \; R = 10^{-3}.$
For each time instant $t$, we define the ambiguity set \eqref{eq:amb_set_e2} by setting $ z_t = 0.5 u_t $  and $ d_t =10^{-10}, $ and we compute the D$^2$-LQG control policy \eqref{eq:policy_e2}
{\rev applying Algorithm 1. This step required 38.25 \si{\second } on a processor Intel\textregistered \hspace{0.01cm}  Core\texttrademark  \hspace{0.01cm}  i7-1355U. } 
{\rev \\
We aim to compare the performance of our approach with three control schemes: the standard LQG controller, the DRC controller with a single relative-entropy constraint (D-LQG) \cite{Petersen2000}, and the Wasserstein DRC (WDRC) controller for linear quadratic problems proposed in \cite{kim2020minimax}. 
In \cite{kim2020minimax}, the deviation between probability distributions is measured in terms of the Wasserstein metric.  Additionally, rather than employing an ambiguity set, the method in \cite{kim2020minimax} pursues distributional robustness by incorporating a penalty term in the objective function. The relevance of this penalty term is controlled by a penalty parameter, which we set to 50 in the following simulations.}\\ 
The control schemes are evaluated by comparing their performance when the noise process ${v}_t $ is generated according to the Linear Fractional Transformation (LFT) model illustrated in Figure \ref{fig:LFT}. 
We recall that LFT model \cite{Petersen2014RobustCO} is frequently used  in modern control theory to represent various sources of uncertainty.
It consists in separating the nominal model from the uncertainty in a feedback interconnection as shown in Figure \ref{fig:LFT}.
The uncertainty operator $\Delta$ is a quantity which is typically unknown but bounded in magnitude.  By suitably selecting the admissible structure of $\Delta$ and the way in which $\Delta$ is bounded in magnitude, we can effectively model the uncertainty on the system dynamics. 
In this example, we assume that $ \bar v_t $ is a white Gaussian noise such that  $\bar v_t \sim f_t,$ and $\Delta$ is a time-invariant matrix given by 
$ \Delta = c [1 \;  1 \; 1], $
with $ c $ being a scalar ranging in the interval $[-1/ \sqrt{3} , 1/ \sqrt{3}].$ 
By using Lemma \ref{lemma:REgass} and the inequality  $\Vert \Delta \Vert^2 \leq I,$ we have that 
$ {\rev \Rc(\tilde f_t || f_t ) } = 
\frac{1}{2}  z_t \tp \Delta \tp \Delta  z_t \leq  \frac{1}{2} \Vert  z_t \Vert ^2.$
In other words, $\tilde f_t \in \Bc_t(x_t, u_t) $ for any $t.$ 
 \\
A plot of the expected closed-loop cost \eqref{eq:cost} versus the uncertainty parameter $\Delta$ for the  controllers is given in Figure \ref{fig:ex1}.
When $\Delta = 0$ the perturbed system corresponds to the nominal one, for which the standard LQG controller is  optimal.
However, the LQG performance rapidly deteriorates when the magnitude of $\Delta$ increases, causing the cost of the system to increase dramatically. 
{\rev Similarly, the WDRC controller  performs well when $\Delta$ is close to zero,  but it fails to ensure robustness for larger values of $\Delta$.} 
It is also evident that the DRC with a single relative entropy constraint \cite{Petersen2000} does not show a satisfactory behavior. 
The explanation lies in the fact that the maximizing player is allowed to allocate most of the mismatch modeling budget to the first time intervals; this leads to extremely pessimistic and conservative conclusions. 
Finally, we notice that the proposed procedure is able to trade off optimality and robustness. Indeed, when the real system coincides with the nominal one the expected closed-loop cost is negligibly larger than the pure LQG optimum, which is clearly a
lower bound for the D$^2$-LQG problem. This cost remains almost constant when the perturbation matrix $\Delta$ is different from zero, giving evidence to the robustness properties of the control system.
This provide a numerical evidence of the fact that, at least in this example, our suboptimal solution is for all practical purposes optimal. This appears to be the case in several simulations 
that we have performed. 
{\rev  
We hasten to remark that we have repeated the previous simulation for different values of the penalty parameter in the WDRC scheme, yielding results consistent with those shown in Figure \ref{fig:ex1}. }
\\
{\rev 
To asses the impact of the initial condition $\bar x_0$ on the performance of the proposed controller, we 
repeat the simulations for different $\bar x_0$ values. 
Specifically, we assume that the real underlying system is generated according to the LFT model of Figure 2 with $\Delta = -0.35 [ 1 \;  1 \; 1],$ and that the initial condition is given by $\bar x_0 = d[1 \; 1 \; 1] $ with $d$ being a scalar ranging in the interval $[-1/\sqrt{3}, 1/\sqrt{3}].$ 
Figure \ref{fig:ex1_cost_vs_x0} summarizes the result, showing the average closed-loop cost as a function of  $\Vert \bar x_0 \Vert.$ 
 The figure demonstrates that the proposed D$^2$-LQG controller consistently outperforms the standard LQG method, the WDRC controller and the D-LQG method across all the tested values of $\bar x_0.$ 
}

\begin{figure} 
		\tikzset{every picture/.style={line width=0.75pt}} 
	\begin{tikzpicture}[x=0.75pt,y=0.75pt,yscale=-0.8,xscale=0.8] 
		
		\draw   (293.72,51) -- (371.39,51) -- (371.39,105.21) -- (293.72,105.21) -- cycle ;
		\draw   (293.72,160.79) -- (371.39,160.79) -- (371.39,215) -- (293.72,215) -- cycle ;
		\draw    (167,201) -- (279,200.88) -- (291,200.98) ;
		\draw [shift={(294,201)}, rotate = 180.44] [fill={rgb, 255:red, 0; green, 0; blue, 0 }  ][line width=0.08]  [draw opacity=0] (8.93,-4.29) -- (0,0) -- (8.93,4.29) -- cycle    ;
		\draw    (371.39,200.09) -- (467,200) ;
		\draw [shift={(470,200)}, rotate = 179.95] [fill={rgb, 255:red, 0; green, 0; blue, 0 }  ][line width=0.08]  [draw opacity=0] (8.93,-4.29) -- (0,0) -- (8.93,4.29) -- cycle    ;
		\draw    (371.39,174.34) -- (436.86,174.34) ;
		\draw    (436.64,87.95) -- (436.75,174.69) ;
		\draw    (436.64,87.95) -- (374.39,88.9) ;
		\draw [shift={(371.39,88.95)}, rotate = 359.12] [fill={rgb, 255:red, 0; green, 0; blue, 0 }  ][line width=0.08]  [draw opacity=0] (8.93,-4.29) -- (0,0) -- (8.93,4.29) -- cycle    ;
		\draw   (216.04,176.37) .. controls (216.04,168.51) and (221.51,162.14) .. (228.25,162.14) .. controls (234.99,162.14) and (240.46,168.51) .. (240.46,176.37) .. controls (240.46,184.23) and (234.99,190.6) .. (228.25,190.6) .. controls (221.51,190.6) and (216.04,184.23) .. (216.04,176.37) -- cycle ;
		\draw    (165,176.37) -- (213.04,176.37) ;
		\draw [shift={(216.04,176.37)}, rotate = 180] [fill={rgb, 255:red, 0; green, 0; blue, 0 }  ][line width=0.08]  [draw opacity=0] (8.93,-4.29) -- (0,0) -- (8.93,4.29) -- cycle    ;
		\draw    (293.61,90.31) -- (228.25,90.31) ;
		\draw    (228.25,90.31) -- (228.25,159.14) ;
		\draw [shift={(228.25,162.14)}, rotate = 270] [fill={rgb, 255:red, 0; green, 0; blue, 0 }  ][line width=0.08]  [draw opacity=0] (8.93,-4.29) -- (0,0) -- (8.93,4.29) -- cycle    ;
		\draw    (240.46,176.37) -- (290.72,177.01) ;
		\draw [shift={(293.72,177.05)}, rotate = 180.73] [fill={rgb, 255:red, 0; green, 0; blue, 0 }  ][line width=0.08]  [draw opacity=0] (8.93,-4.29) -- (0,0) -- (8.93,4.29) -- cycle    ;
		
		\draw (302 ,172) node [anchor=north west][inner sep=0.75pt]   [align=center] {Nominal\\System};
		\draw (320 , 70) node [anchor=north west][inner sep=0.75pt]  [font=\Large]  {$\Delta $};
		\draw (255,  205) node [anchor=north west][inner sep=0.75pt]    {$u_{t}$};
		\draw (438,  205) node [anchor=north west][inner sep=0.75pt]    {$x_{t}$};
		\draw (400, 73) node [anchor=north west][inner sep=0.75pt]    {$ z_{t}$};
		\draw (238, 150) node [anchor=north west][inner sep=0.75pt]    [align=center] {${v}_{t} \sim \tilde f_t$};
		\draw (162, 150) node [anchor=north west][inner sep=0.75pt]    [align=center] {$\bar v_{t} \sim f_t$};		
	\end{tikzpicture}
	 \caption{LTF uncertain model.} \label{fig:LFT}
\end{figure}

\begin{figure} 
{\rev 	
\centering
	\includegraphics[width= \linewidth]{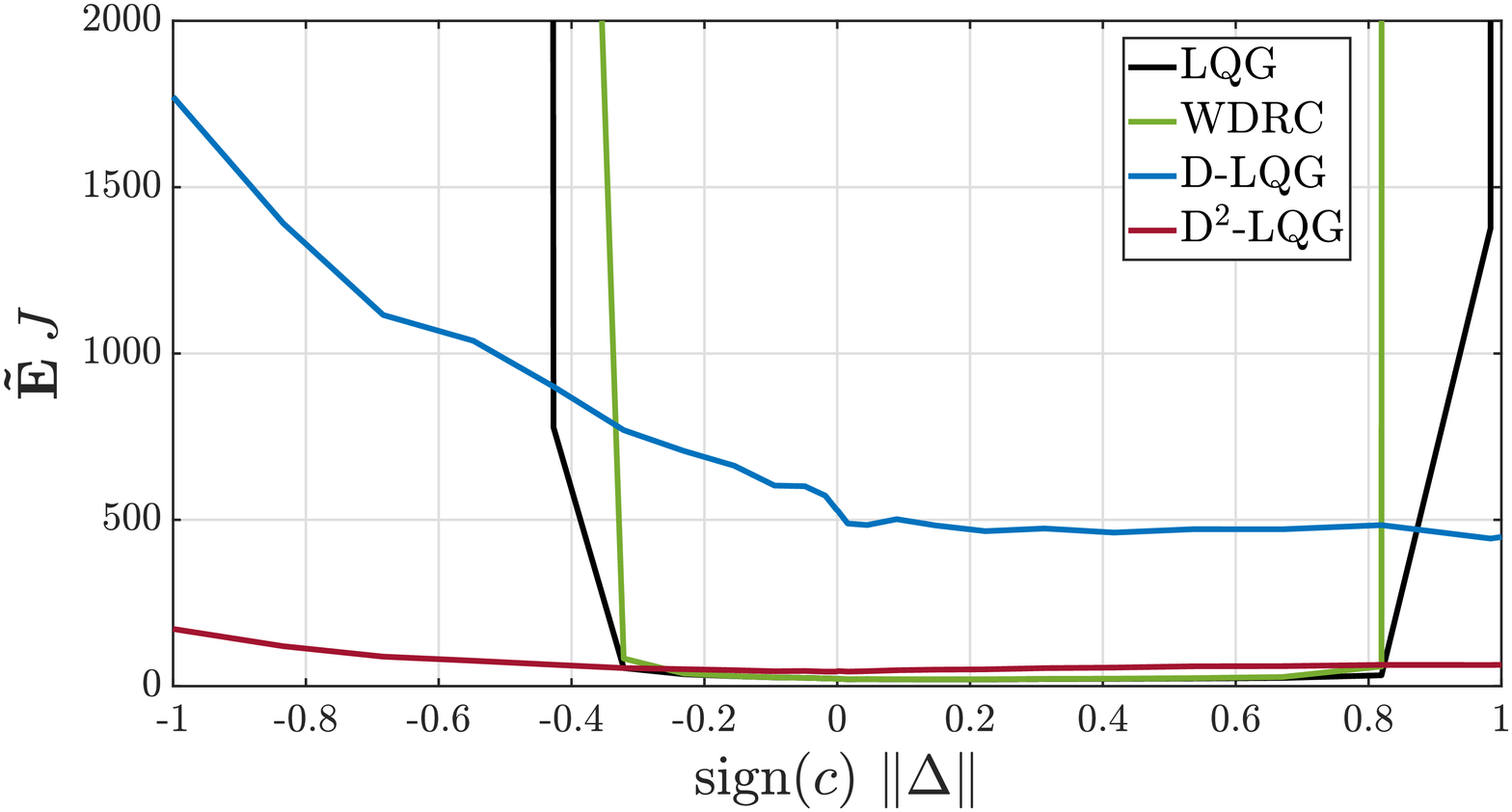}
	\caption{ Expected closed-loop cost versus the uncertain parameter $\Delta$, with $\Delta = c[1 \; 1 \; 1]$, $c \in [-1/\sqrt{3}, 1/\sqrt{3}].$ We compare the performance of the standard LQG controller (in black), the WDRC controller (in green), the D-LQG with a single-constraint (in blue) and the proposed  D$^2$-LQG controller (in red). }
	\label{fig:ex1}
 }
\end{figure}

\begin{figure} 
{\rev 
	\centering
	\includegraphics[width=\linewidth]{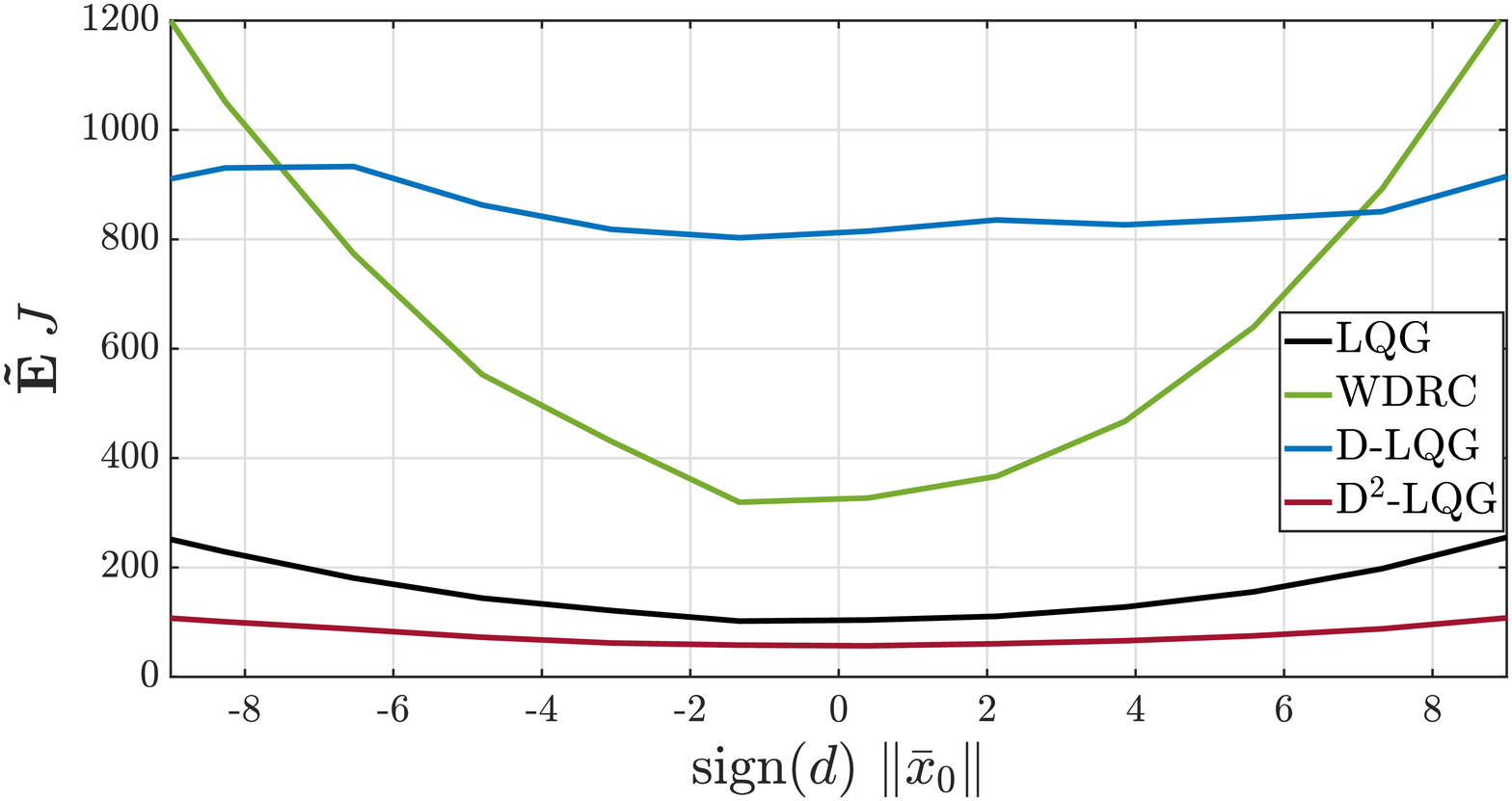}
	\caption{ Expected closed-loop cost versus the initial condition  $\bar x_0$, with $\bar x_0 = d[1 \; 1 \; 1]$, $d \in [-1/\sqrt{3}, 1/\sqrt{3}]$. We compare the performance of the standard LQG controller (in black), the WDRC controller (in green), the D-LQG with a single-constraint (in blue) and the proposed  D$^2$-LQG controller (in red). }
	\label{fig:ex1_cost_vs_x0}
 }
\end{figure}

\section{Conclusions}\label{sec:conc}
We have proposed a new LQG problem and  derived a control policy which is distributionally robust under distributed uncertainty.  As discussed in detail in the introduction, there are strong motivations for considering a distributed constraint on the uncertainties. For this reason, several attempts have been appeared in the recent literature addressing this kind of problem. To the best of our knowledge this is the first time in the literature that this kind of D$^2$-LQG problems have been solved without relaxations or approximations.
Our simulation experiments suggest the following conclusions:
\begin{enumerate}
\item
The standard LQG 
solution is quite fragile. In fact, the  LQG optimal control policy applied to a system 
featuring modest perturbations with respect to the nominal model may
yield an unstable closed-loop system. Both our method and the method proposed in 
\cite{Petersen2000} are, on the other hand, resilient to perturbations.
\item 
Our method to deal with uncertainties appears to be much less conservative than that in 
\cite{Petersen2000}. In fact, our performance are not significantly different from the optimal LQG ones for the nominal model while the performance index obtained by using the method in \cite{Petersen2000} is much larger. Moreover, 
even when the difference between the nominal and the actual model is significant, robustness against instability is guaranteed by both methods but in our method the value of the performance index is significantly smaller.
\end{enumerate}

{\rev Extending the results to the output feedback framework is an interesting line of future research. Additionally, integrating adaptive control schemes with the proposed DRC framework could be an area for future work, potentially combining the robustness of DRC with the flexibility of adaptive control.
}

\bibliographystyle{plain}        
\bibliography{biblio_LQGrobusto}           



\appendix

\section{Relative Entropy and Optimization}\label{sec:app_KL}
In this section, we briefly recall the definition and some well-known properties of the relative entropy (see, for example, \cite{Daipra1996}). 
Given two probability measures $\tilde f$ and $f$ defined on $\R^n$, we measure the deviation  of $\tilde f $  from $f$ in terms of the \emph{relative entropy} (or \emph{Kullback-Leibler} divergence)
$$
{\rev \Rc(\tilde f || f)} = 
\int_{\R^n} \ln \frac{\tilde f(x)}{f(x)} \tilde f(x) dx.
$$
${\rev \Rc(\cdot || \cdot)}$ is not a distance since it is not symmetric, and, more importantly, it does not satisfy the triangle inequality. 
However it is a pseudo-distance as it satisfies the following properties:
$\rev \Rc(\tilde f|| f) \geq 0$ and $\rev \Rc(\tilde f|| f) = 0$ if and only if $ \tilde f(x) = f(x)$ a.e.
Moreover, 	it is well known that, for a given probability density function $f$, $\rev \Rc(\tilde f|| f) $ is a strictly convex function of $ \tilde f$ on the set of probability density functions. 

\begin{lemma} \label{lemma:REgass}
Suppose that $\tilde f$ and $f$ are Gaussian density functions on $\R^n$ with the same (non-singular) covariance matrix $V$: $ f \sim \Nc (v, V)$ and $ \tilde f \sim \Nc (\tilde v, V) $
Then
$
{\rev \Rc(\tilde f || f)} = 
\frac{1}{2} (\tilde v - v) \tp V\inv (\tilde v - v).$
\end{lemma}

Finally, we recall the duality between free energy and relative entropy which underlies the solution to the proposed control problem.
\begin{lemma}\label{lemma:KLopt}
For a given probability density function $f:\R^n \to \R_+$ and a measurable function $J : \R^n \to \R$ bounded from below,  then 
\beq \label{eq:daipra}
\sup_{\tilde f(x)} \! \int \! J(x) \tilde f(x) dx - {\rev \Rc(\tilde f|| f)} = \log \!\int \! e^{J(x)} f(x) dx
\eeq
where the supremum is taken over the set of all the probability density functions $\tilde f(x)$ on $\R^n$ such that
$ \rev{ \Rc(\tilde f || f) }< \infty. $ 
Moreover, if $ \int J(x) e^{J(x)} f(x) dx < \infty, $ then the supremum in \eqref{eq:daipra} is achieved by the probability density function 
\beq \label{eq:least_fav}
\tilde f^\star =  \frac{ e^{J}  f}{ \int e^{J(\bar x)} f(\bar x) d \bar x}.
\eeq
\end{lemma}

\section{Convex Optimization}\label{sec:app_opt}
In this section we collect some relevant properties on optimization and convexity underlying the solution the proposed control problem.

\begin{lemma}\label{lemma:convex} 	\cite[p.81]{boyd:vandenberghe:2004}
	Let $X$ be a normed space. Let $\{f_{\alpha} (x) | \alpha \in I \}$ be a collection of functions with the same domain $K.$ If $K$ is  a convex subset of $X$ and $f_{\alpha} (x) $ is  a convex function for each $\alpha$, then 	$	g(x) := \sup_{\alpha \in I} \; f_{\alpha}(x) 	$
	is also convex.
\end{lemma}

\begin{lemma}\label{lemma:convex_min}	\cite[p.87]{boyd:vandenberghe:2004}
	Let $f(x,y)$  be convex in $(x,y)$ and $K$  be a convex non-empty set. Then the function 
	$
	g(x) = \inf_{y \in K} f(x,y)
	$
	is convex in $x$ provided that $g(x) > -\infty $ for some $x.$
\end{lemma}

\begin{theorem}\cite[Corollary 3.3]{sion1958general} \label{lemma:minmax}	
Let $X$ and $Y$ convex subsets of a linear topological space, one of which is compact. 
If $ f : X \times Y \to \R$ with 
$f(\cdot, y)$ upper semicontinuous and quasi-concave on $X$ for any $y \in Y $,  and 
$f(x, \cdot)$ lower semicontinuous and quasi-convex on $Y$ for any $x \in X,$ then 
$$
\sup_{x \in X}\;  \inf_{y \in Y} f(x,y) =\sup_{y\in Y} \; \inf_{x \in X} f(x,y).
$$
\end{theorem}

\begin{theorem}\label{lemma:luenberger} \cite[pag. 224-225]{luenberger}
	Let $X$ be a linear vector space and $\Omega$ be a convex subset of $X$. Let $f$ be a real-valued concave function on $\Omega$ and $g: X \to \R$ a convex map. Suppose there exists  $x_1 \in \Omega$ such that $ g(x_1) < 0.$
	Let
	\beq \label{eq:dual1}
	\mu_0 = \sup_{ \substack{x \in \Omega \\ g(x) \leq 0}} f(x)
	\eeq
	and assume $\mu_0$ is finite. Then, 
	\beq \label{eq:strongduality}
	\sup_{ \substack{x \in \Omega \\ g(x) \leq 0 }} f(x) = \min_{\tau \geq 0} \; \sup_{x \in \Omega} [f(x) - \tau g(x)]
	\eeq 
	and the minimum in the right side is achieved for $\tau^o \geq 0.$ 	
\end{theorem}

\section{Integral of multivariate exponential functions}\label{sec:app_gauss}
\begin{lemma}\label{lemma:GaussIntegral} \cite{Zee2010}
Given  $\Sigma \in \Sb^n$, $x, b \in \R^n$ and $c \in \R$
$$
\int_{\R^n} e^{- \frac{1}{2}x\tp \Sigma x + b\tp x + c} dx= \sqrt{\frac{(2\pi)^n}{|\Sigma|}} e^{\frac{b\tp \Sigma^{-1} b}{2} + c}
$$
if and only if $ \Sigma \in \Sb_{++}^n$. Otherwise the integral diverges. 
\end{lemma}

\begin{lemma} \label{lemm:worstpdf}
	Let $f(x) = \Nc(0, \Sigma)$ with $\Sigma \in \Sb_{++}^n $, and $J(x) = \frac{1}{2} x\tp G x + b\tp x$ with $ b \in \R^n$ and $G\in \Sb_{+}^n.$
	Assume that $(\Sigma\inv - G ) \in \Sb^n_{++}.$ Then 
    $$
	\tilde f(x) = \frac{  e^{J(x)}f(x) }{ \int_{\R^n} e^{J(\bar x)} f(\bar x) d \bar x} 
    $$
	is a Gaussian density function $
	 \tilde f(x) = \Nc \big( (\Sigma \inv - G)\inv b ,  (\Sigma \inv - G)\inv  \big) .$
\end{lemma}
\proof
Exploiting Lemma \ref{lemma:GaussIntegral} we have that
\begin{align*}
\int e^{J(x)} f(x) dx  
&=  
\frac{1}{\sqrt{(2\pi)^n |\Sigma|}}
\int e^{-\frac{1}{2} x\tp (\Sigma \inv - G) x  + b\tp x} dx 
\\
& = \frac{1}{\sqrt{ |I - \Sigma G)| }} 
e^{\frac{1}{2} b\tp (\Sigma \inv - G)\inv b} .
\end{align*}
Moreover, by completing the square, we can see that 
$$e^{J(x)} f(x)  = 
\frac{1}{\sqrt{(2 \pi)^n |\Sigma|}} 
e^{-\frac{1}{2} (x - \tilde \mu)\tp \tilde \Sigma \inv (x - \tilde \mu) + b\tp \tilde \Sigma\inv b} $$ 
where $ \tilde \mu =(\Sigma \inv - G)\inv b $ and $\tilde \Sigma = (\Sigma \inv - G)\inv . $
Now, the thesis follows immediately. $\hfill \qed$

\end{document}